\documentclass[12pt,letterpaper]{article}
\usepackage[utf8]{inputenc}
\usepackage{amsmath,amsthm,amssymb,amsfonts}

\usepackage{graphicx}
\usepackage{epstopdf}
\usepackage{epsfig}

\usepackage{microtype}
\DisableLigatures[f]{encoding = *, family = * }
\usepackage{rotating}
\usepackage[aboveskip=1pt,labelfont=bf,labelsep=period,justification=justified,singlelinecheck=off]{caption}

\usepackage[font=small,labelfont=bf,labelsep=space]{caption}
\usepackage{cite}

\makeatletter
\renewcommand{\@biblabel}[1]{\quad#1.}
\makeatother

\date{}

\begin{document}

\begin{flushleft}
{\Large
\textbf{Inferring Synaptic Structure in presence of Neural Interaction Time Scales}
}\\

\vspace{5mm}
\textbf{Cristiano Capone$^{1,2,5,\ast}$,
Carla Filosa$^{1}$,
Guido Gigante$^{2,3}$,}\\
\textbf{Federico Ricci-Tersenghi$^{1,4,5}$,
Paolo Del Giudice$^{2,5}$}\\
\vspace{5mm}
{\small
{\bf 1} Physics Department, Sapienza University, Rome, Italy \\
{\bf 2} Italian Institute of Health, Rome, Italy\\
{\bf 3} Mperience s.r.l., Rome, Italy\\
{\bf 4} IPCF-CNR, UOS Roma, Italy\\
{\bf 5} INFN, Sezione di Roma 1, Italy\\
$\ast$ E-mail: cristiano.capone@iss.infn.it}
\end{flushleft}

\section*{Abstract}

Biological networks display a variety of activity patterns reflecting a web of interactions that is complex both in space and time. Yet inference methods have mainly focused on reconstructing, from the network's activity, the spatial structure, by assuming equilibrium conditions or, more recently, a probabilistic dynamics with a single arbitrary time-step. Here we show that, under this latter assumption, the inference procedure fails to reconstruct the synaptic matrix of a network of integrate-and-fire neurons when the chosen time scale of interaction does not closely match the synaptic delay or when no single time scale for the interaction can be identified; such failure, moreover, exposes a distinctive bias of the inference method that can lead to infer as inhibitory the excitatory synapses with interaction time scales longer than the model's time-step. We therefore introduce a new two-step method, that first infers through cross-correlation profiles the delay-structure of the network and then reconstructs the synaptic matrix, and successfully test it on networks with different topologies and in different activity regimes. Although step one is able to accurately recover the delay-structure of the network, thus getting rid of any \textit{a priori} guess about the time scales of the interaction, the inference method introduces nonetheless an arbitrary time scale, the time-bin $dt$ used to binarize the spike trains. We therefore analytically and numerically study how the choice of $dt$ affects the inference in our network model, finding that the relationship between the inferred couplings and the real synaptic efficacies, albeit being quadratic in both cases, depends critically on $dt$ for the excitatory synapses only, whilst being basically independent of it for the inhibitory ones.

\newpage
\section*{Introduction}

The attempt to infer synaptic connectivity from correlations between neural activities has a long history (see, \textit{e.g.}, \cite{eggermont1990correlative} and references therein; for recent developments, see, \textit{e.g.}, \cite{pernice2013reconstruction}). In this context, a long recognized problem is that, since the real neuronal network is dramatically under-sampled by electrophysiology experiments, one cannot remove ambiguities as to whether observed correlations depend on direct synaptic connections or on indirect loops through unobserved components of the network. The advent of multi-electrode arrays bringing the number of simultaneously recordings to several tens (both \textit{in-vitro}, e.g. MEA, ad \textit{in-vivo}, e.g. Utah arrays) does not resolve the issue, since still a tiny fraction of the neural population can be sampled; however, it does offer an option for the reconstruction of some forms of effective synaptic connectivities and opens the possibility to address questions which were previously out of reach.

For instance, in the seminal work by Schneidman \textit{et al.} \cite{bialek2006weak}, based on the so-called Inverse Ising inference method which is the substrate of the present work, it was possible to assess the share of network information accounted for by pairwise correlations. Such models were based on maximum entropy estimates which, under the assumptions of pairwise interactions, provide couplings and external fields for the Gibbs equilibrium distribution of an equivalent Ising model, and do not possess any inherent time scale. The obvious interest in relaxing the assumption of equilibrium for modeling neural data, led to the development of inference methods based on a kinetic formulation of the equivalent Ising system, that results in maximum-likelihood estimations of the transition probabilities between subsequent states of the system; in this way, one can account for non-stationary neural data, and non-symmetric synaptic couplings \cite{marre2009prediction,hertz2011ising,roudi2011mean}. 
Compared to other methods to establish statistical models of multiple recordings, such inference methods, rooted in analogies with the statistical physics of complex systems, claim to afford an easier link with biologically meaningful quantities \cite{bialek2012biophysics,hertz2011ising}. We also remark that Kinetic Ising inference methods can be seen as a special case of Generalized Linear Models \cite{truccolo2005point ,pillow2008spatio,roudi2015multi}, with a one-step time kernel.

We consider networks of spiking, integrate-and-fire neurons, sparsely coupled through excitatory and inhibitory synapses. The sampled spike trains were binarized by choosing a time-bin such that two or more spikes fell in the same time-bin with negligible probability. This we regard as a minimal requirement, not to loose information about correlations at the single spike level. Beyond this requirement, we study the impact of the chosen time-bin on the quality of the inference procedure to estimate synaptic couplings, by checking the results of an established method based on the Kinetic Ising Model, for a network of spiking neurons, and illustrate its limitations. We therefore introduce a two-step method to first estimate, for each sampled neurons pair, a characteristic time scale of their interaction (spike transmission), and then to use such estimated time scales as time-lags in the modified Kinetic Ising Model we propose. Finally we estimate analytically, and verify numerically, the relationship between true and inferred synapses, and its dependence on the time-bin $dt$ chosen to binarize the data.

\section*{Results}
We simulated sparsely coupled networks of spiking, integrate-and-fire neurons, interacting through excitatory and inhibitory synapses. The sampled spike trains were binarized by choosing a time-bin $dt$ such that the probability two or more spikes falling in the same time-bin was negligible; apart from this requirement, at this stage the choice of the time-bin was arbitrary. We denote the spike train of neuron $i$ by $S_i(t)$, where $S_i(t) = 1$ if a spike is recorded in bin $t$ and $S_i(t) = 0$ otherwise. 
We assume that the data have been generated by the Ising model evolving in accordance with the Glauber dynamics \cite{glauber1963time, marre2009prediction, roudi2011mean}, so that, at each time-step, $S_i(t + dt)$ is sampled according to the probability distribution:
\begin{equation}
\label{eq.glauberDynamics}
P(S_{i}(t + dt)|\mathbf{S}(t))=\frac{\exp{\left[S_{i}(t + dt)H_{i}(t)\right]}}{1+\exp{H_{i}(t)}} \quad \mathrm{with} \; H_{i}(t)=h_{i}+\sum_{j}J_{ij}S_{j}(t)
\end{equation}
that depends on the total ``field'' $H_i(t)$ felt by neuron $i$, and generated by all the neurons in the network through the synaptic matrix $J$. Being $H_i(t)$ a function of the state of the system at time $t$ only, the dynamics in Eq.~(\ref{eq.glauberDynamics}) is Markovian. To infer the best parameters $J$ and $h$, following \cite{roudi2011mean} (see Methods), we resorted to a mean-field approximation of the gradient $\partial L / \partial J_{ij}$ of the likelihood of the data being generated by the model; under this approximation, the equations $\partial L / \partial J_{ij} = 0$ are linear in $J$ and are then easily invertible. It will be shown in Methods that the following relation holds:
\begin{equation}
\label{eq.jInferApprox1}
J_{Inf,ij} \simeq \frac{\langle S_i | S_j = 1 \rangle - m_i}{m_i \, (1 - m_i) \, (1 - m_j)}
\end{equation}
where $m_i{} = \langle S_i \rangle $ and $\langle S_i | S_j = 1 \rangle$ denotes the conditional probability that neuron $i{}$ fires in the same time-bin $dt$ in which it receives a spike from pre-synaptic neuron $j$.

\subsection*{Role of the time-bin for the quality of inference}
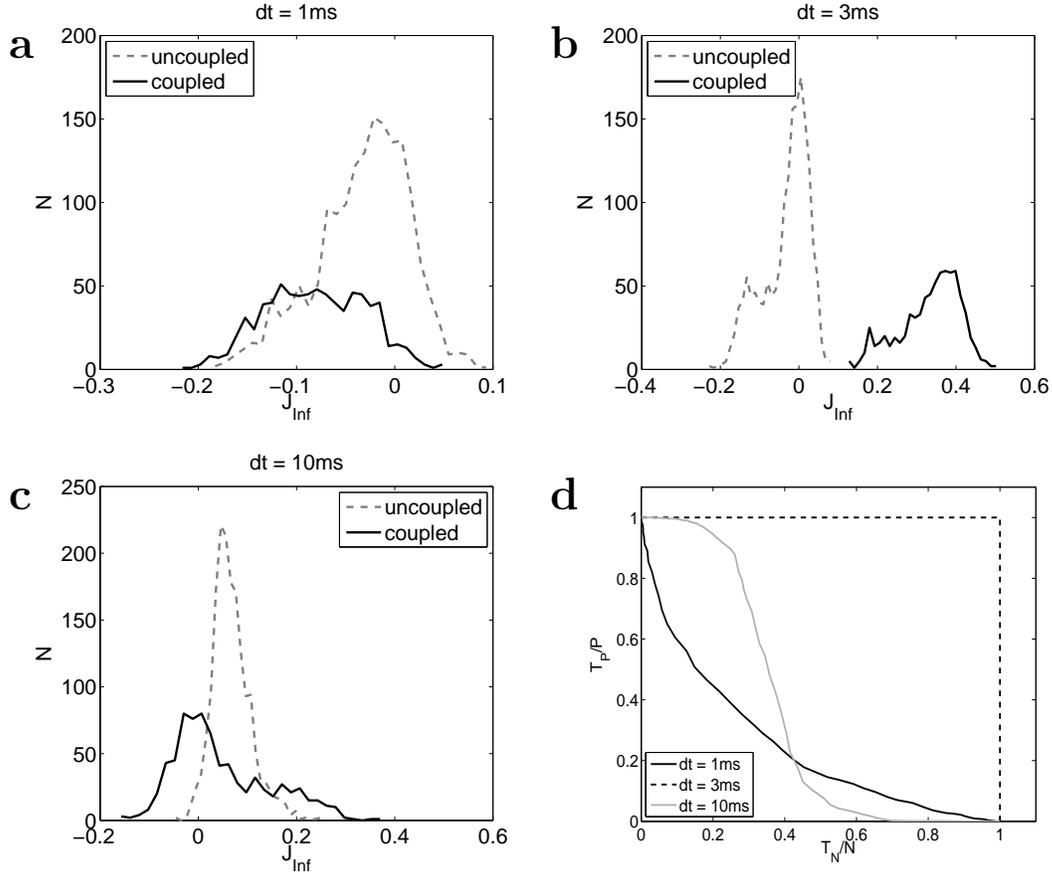
\begin{figure}[h!]
\begin{center}
\setlength{\unitlength}{\textwidth}
\begin{picture}(1,0.75)
  \put(0.04,0.375)
  {
    \epsfig
    {
	file=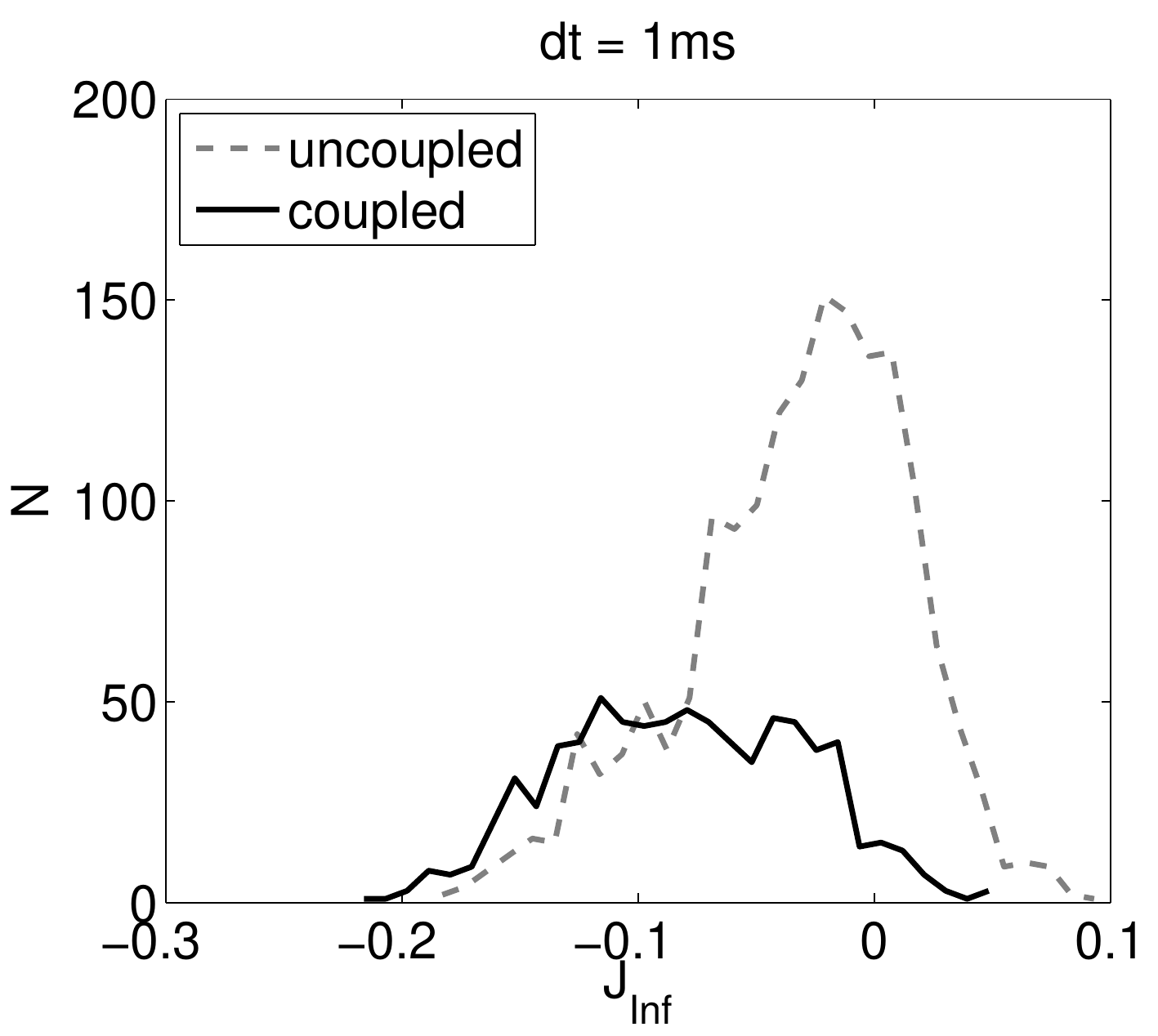,
    width=0.4\unitlength,
    }
  }
  \put(0.49,0.375)
  {
    \epsfig
    {
	file=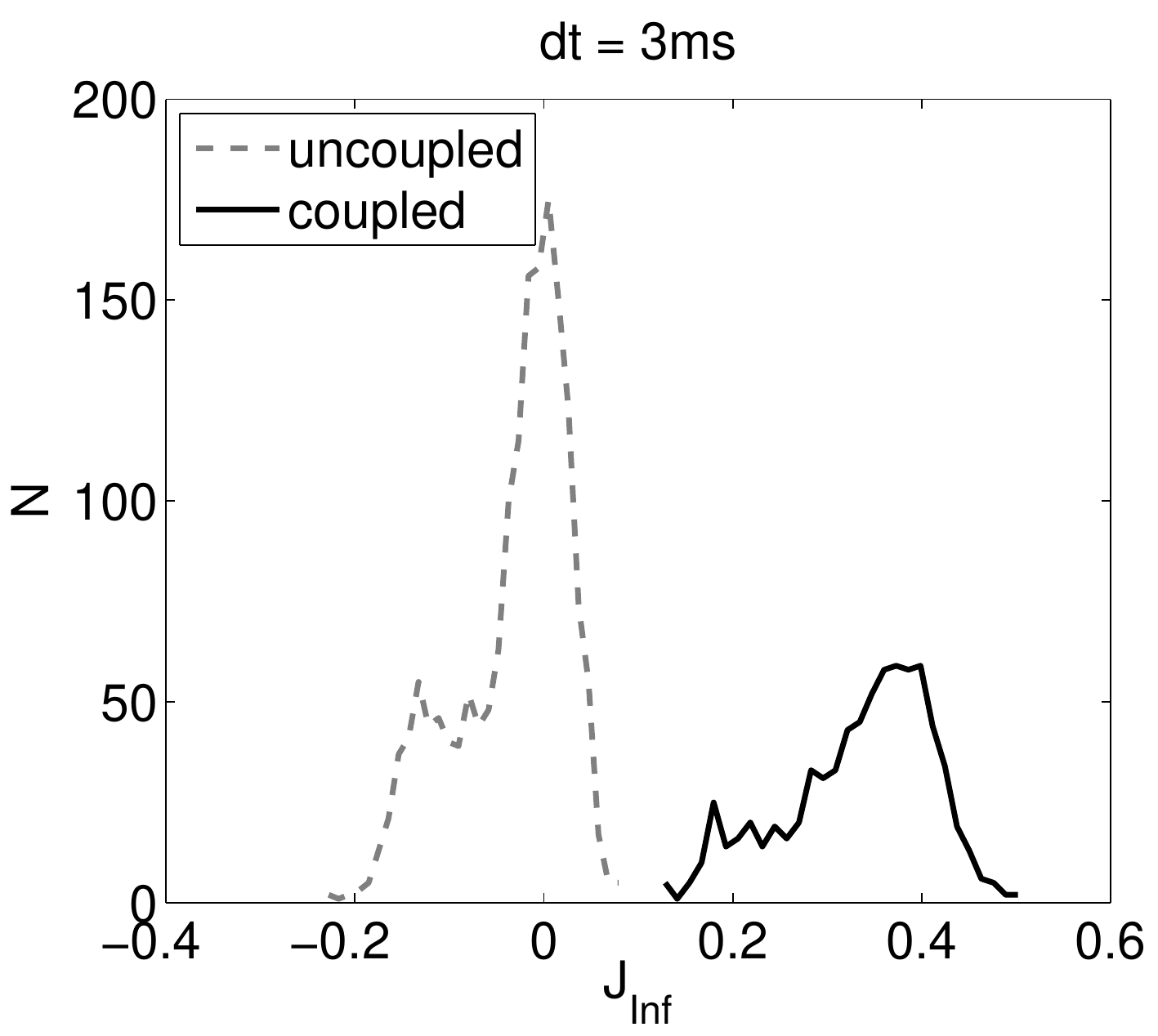,
    width=0.4\unitlength,
    }
  }
  
  \put(0.04,0.)
  {
    \epsfig 
    {
	file=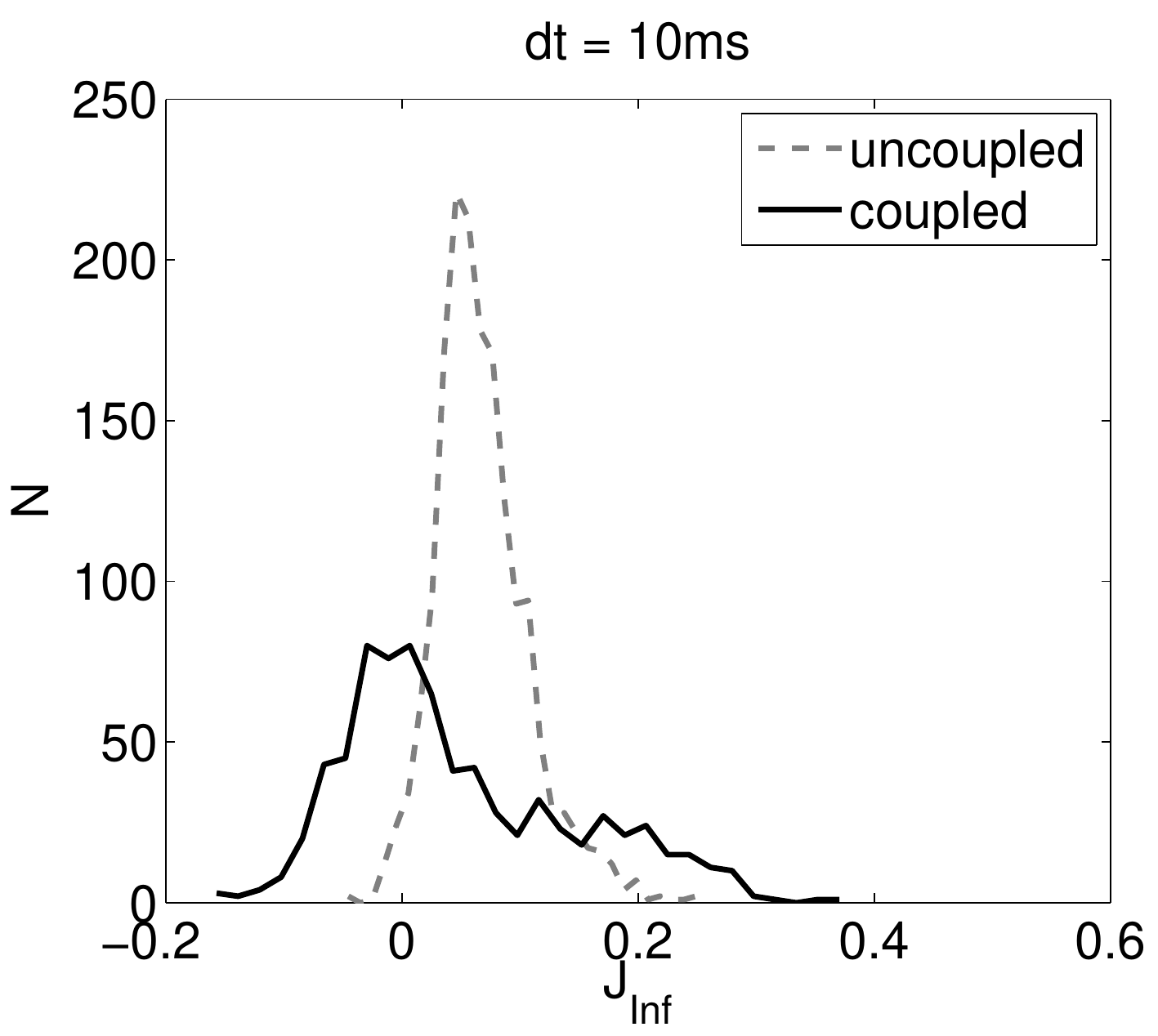,
    width=0.4\unitlength,
    }
  }
  \put(0.5,0.007)
  {
    \epsfig
    {
    file=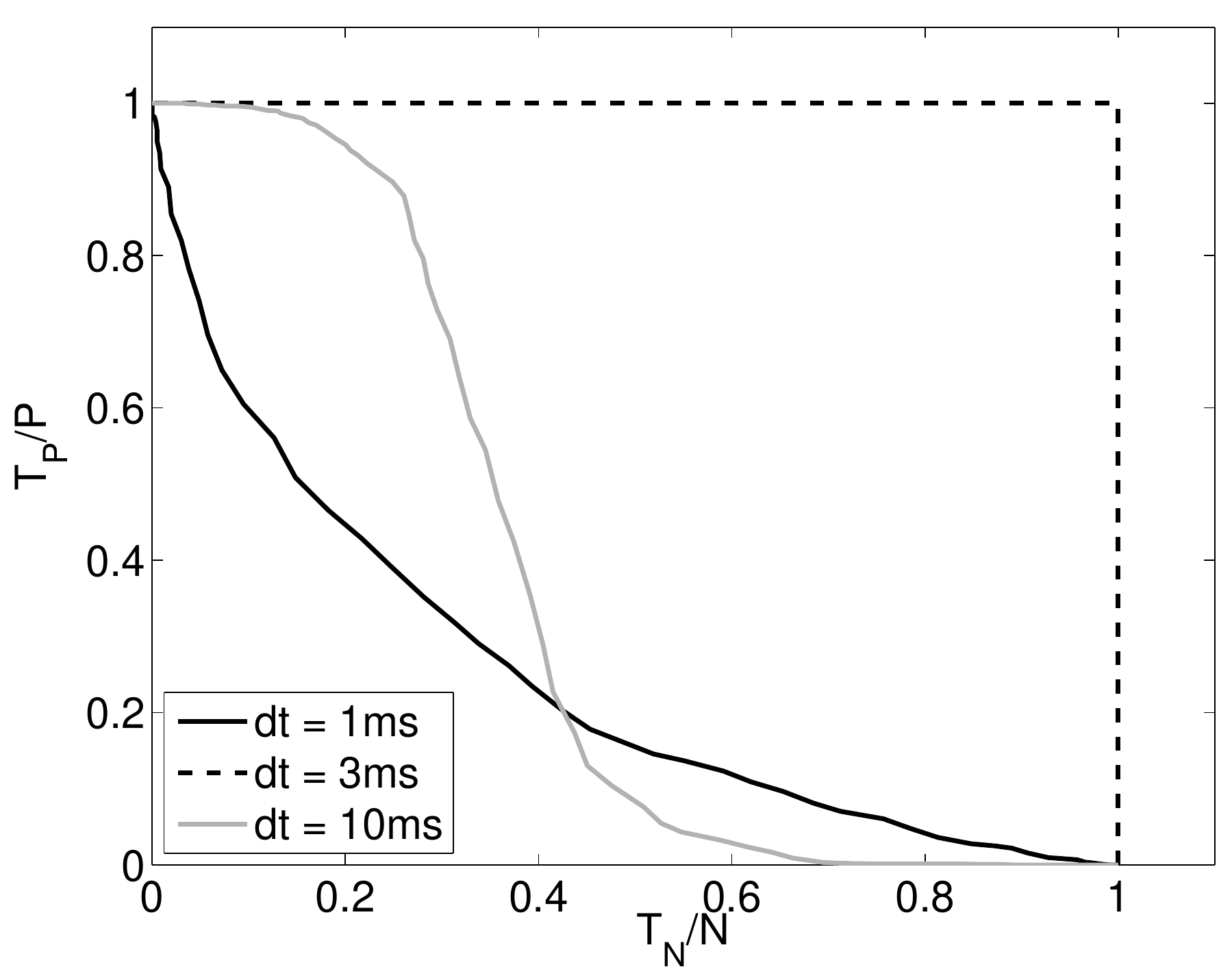,
    width=0.38\unitlength,
    height=0.325\unitlength
    }
  }
  \put(0.03, 0.68){\makebox{\textbf{\Large a}}}
  \put(0.48,0.68){\makebox{\textbf{\Large b}}}
  \put(0.03, 0.305){\makebox{\textbf{\Large c}}}
  \put(0.48, 0.305){\makebox{\textbf{\Large d}}}
\end{picture}
\caption{Inference results for the Kinetic Ising Model with three choices of the time-bin $dt$ ($1,\,3,\,10$ ms), on spike trains generated by a network of $50$ neurons, sparsely connected with probability $c = 0.3$; actual synapses have all equal efficacy $J_{ij} = 0.9$ mV and equal spike transmission delay $\delta = 3$ ms. In panels a-c we plot the histograms of inferred synaptic couplings, for existing synaptic contacts (solid lines) and for those corresponding to unconnected pairs (dashed lines, roughly centered around $0$). The separation between the two histograms is acceptable only for $dt = \delta = 3$ ms (top-right panel); also note, for $dt = 1$ ms, how the solid line peaks around a negative inferred value: the excitatory synapses are in fact inferred as inhibitory -- see text for a discussion of this puzzling result. In panel d the ROC curves corresponding to the three choices of $dt$ are presented; the fraction of correctly identified existing synapses (``true positive'' $T_P / P$) against the fraction of correctly identified unconnected pairs (``true negative'' $T_N / N$) is plotted for moving discrimination threshold: the dashed line, corresponding to $dt = 3$ ms, clearly surmounts the other two, to all effects allowing for a perfect discrimination. Neurons fire at about $50$ Hz; the total recording length is $500$ s.}
\label{figure1}
\end{center}
\end{figure}

In Fig.~\ref{figure1}a-c, we show the inference results for three choices of the time-bin $dt$ ($1,\,3,\,10$ ms), for a (purely excitatory) network of $N = 50$ neurons with equal spike transmission delay $\delta = 3$ ms for all connected neuron pairs. In each panel we plot the histograms of inferred synapses, separately for those corresponding to actually existing synaptic contacts ($30\%$ out of $N\,(N-1)$), and those corresponding to unconnected pairs. It is seen that the expected separation between the peak around zero (corresponding to unconnected pairs) and the histogram corresponding to existing synapses is acceptable only for $dt = 3$ ms, that is when the time-bin $dt$ equals the delay $\delta{}$.

\begin{figure}[h!]
\begin{center}
\setlength{\unitlength}{\textwidth}
\begin{picture}(1,0.75)
  \put(0.04,0.375)
  {
    \epsfig
    {
	file=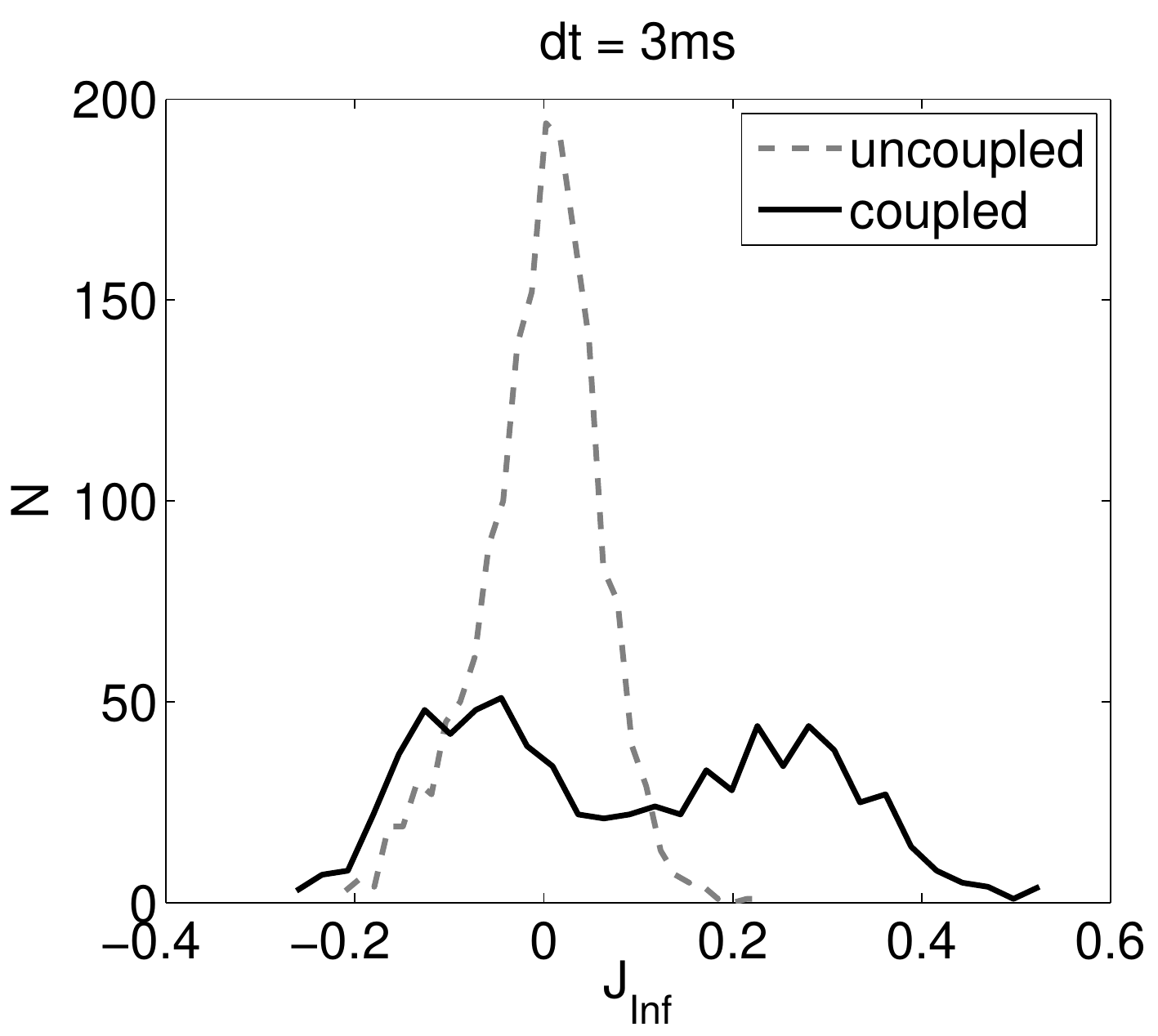,
    width=0.4\unitlength,
    }
  }
  \put(0.49,0.375)
  {
    \epsfig
    {
	file=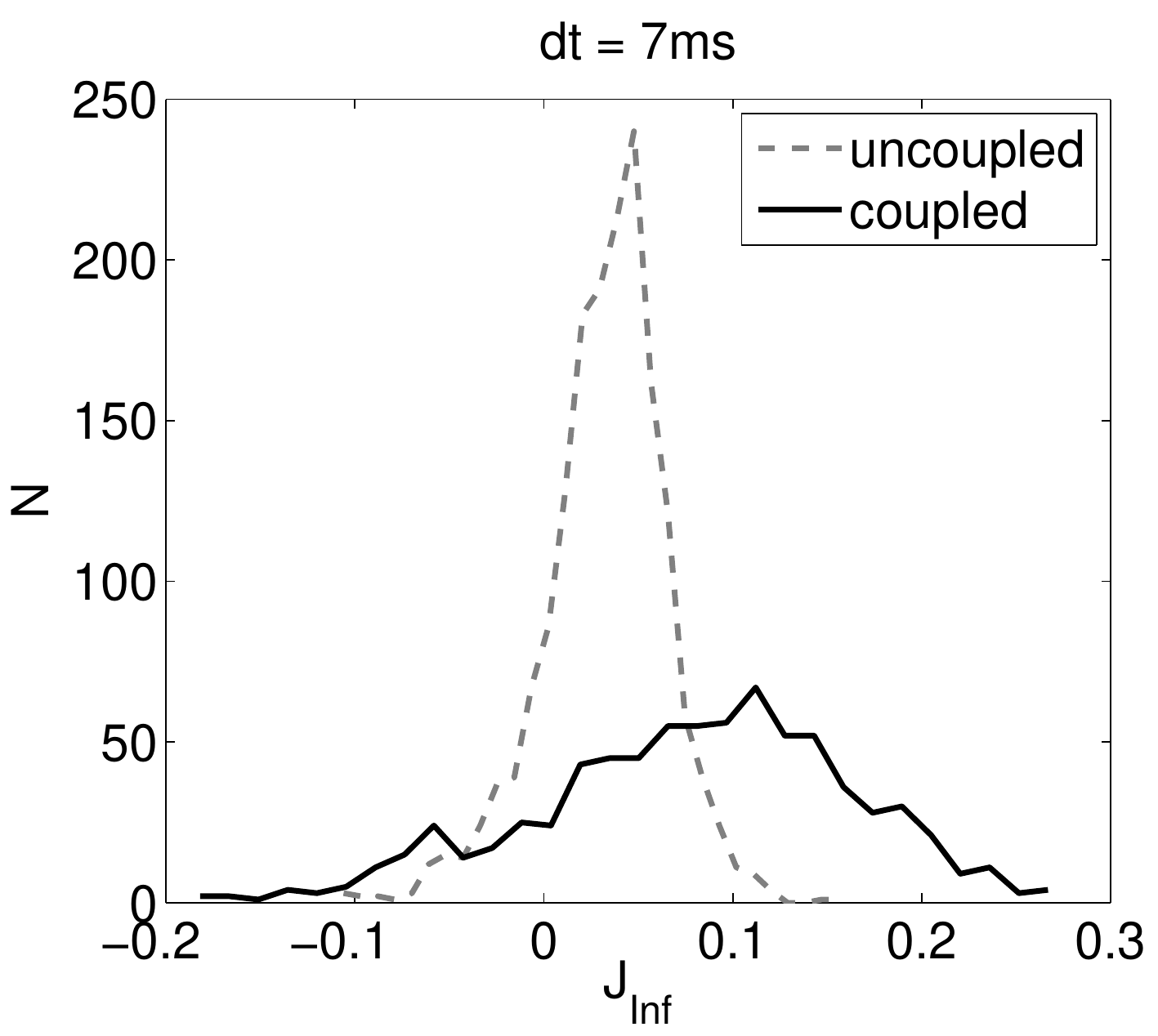,
    width=0.4\unitlength,
    }
  }
  
  \put(0.04,0.)
  {
    \epsfig 
    {
	file=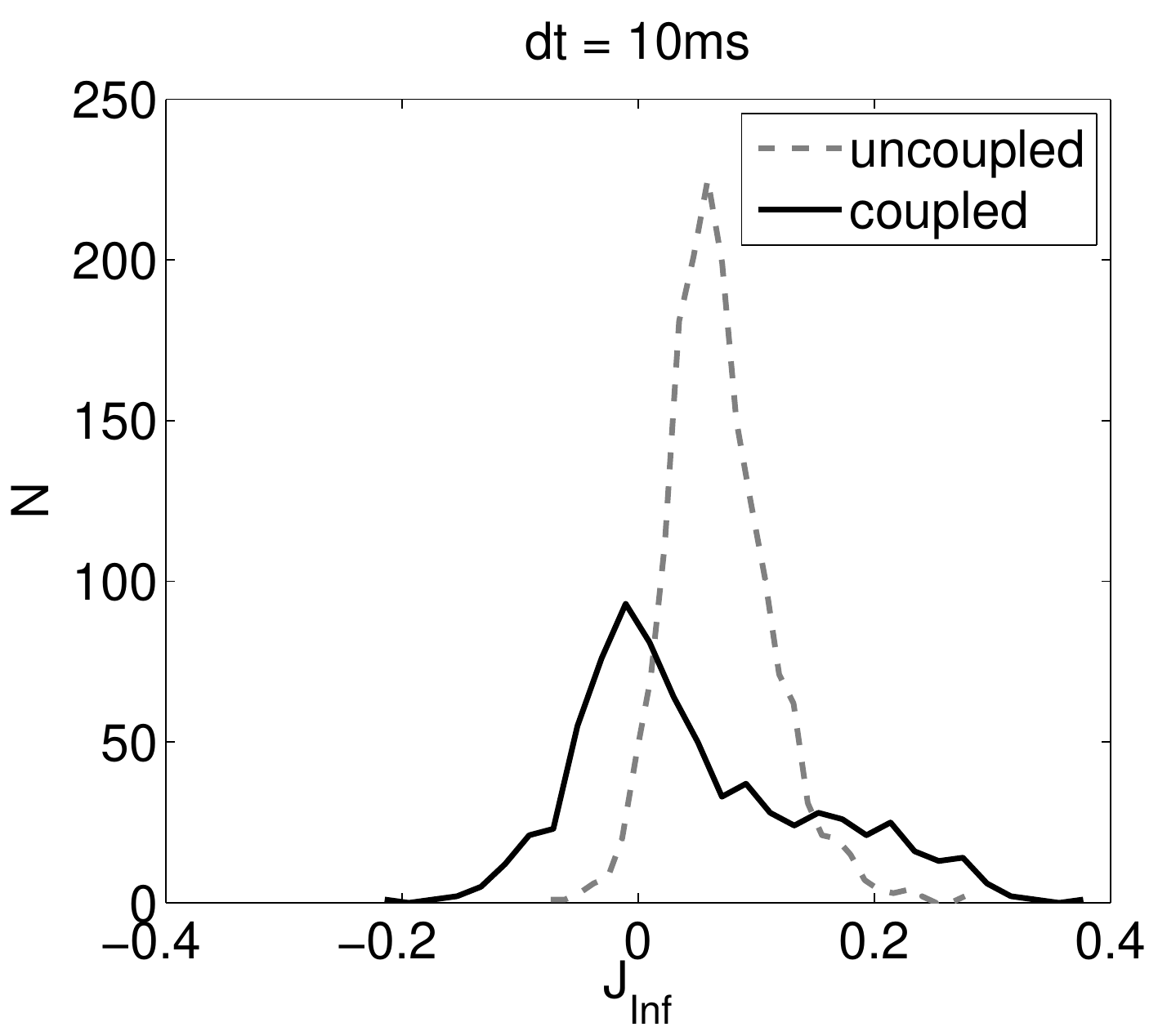,
    width=0.4\unitlength,
    }
  }
  \put(0.475,0.)
  {
    \epsfig
    {
    file=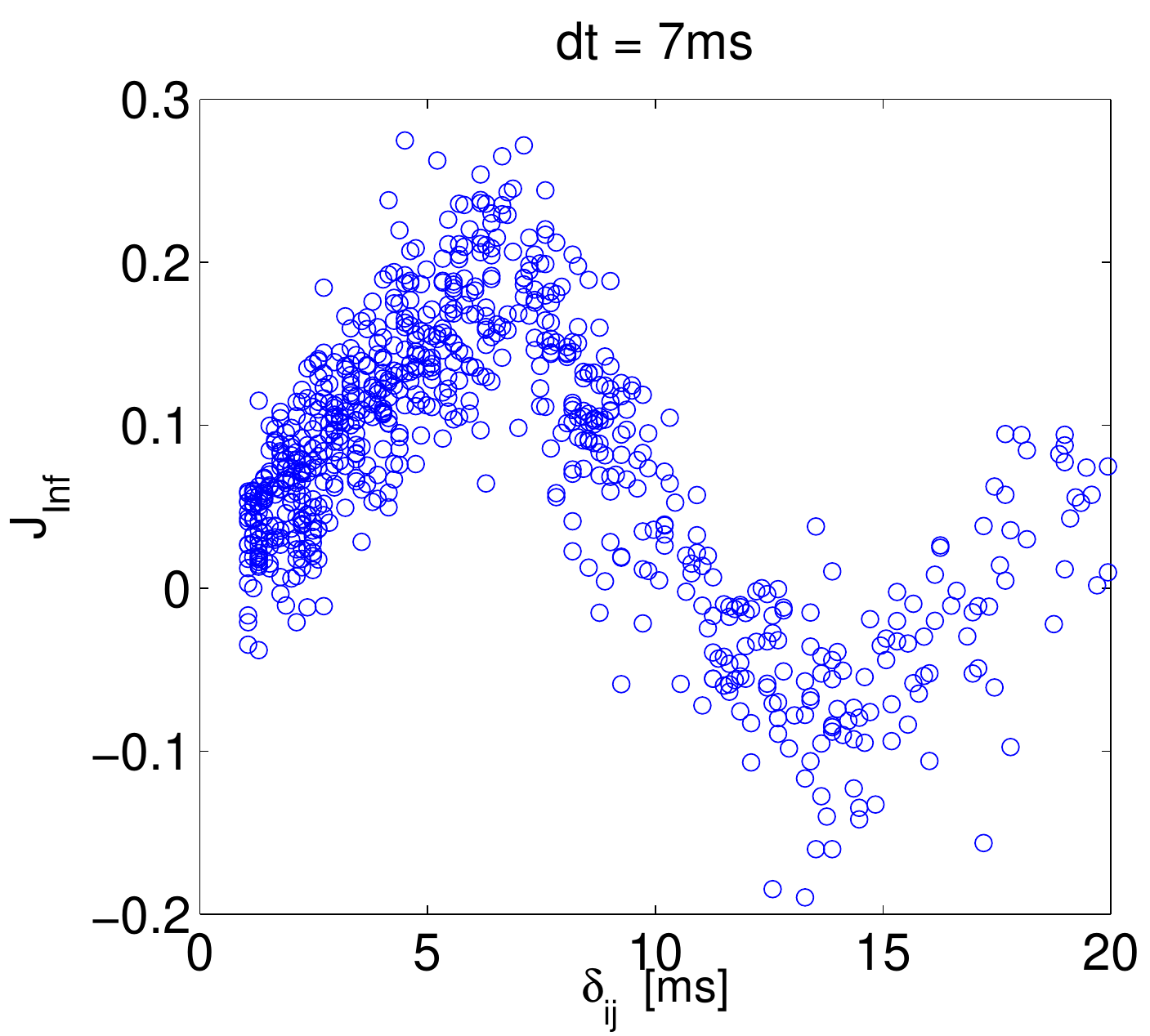,
    width=0.4178\unitlength,
    height=0.36\unitlength
    }
  }
  \put(0.115, 0.665){\makebox{\textbf{\Large a}}}
  \put(0.565,0.665){\makebox{\textbf{\Large b}}}
  \put(0.115, 0.29){\makebox{\textbf{\Large c}}}
  \put(0.565, 0.29){\makebox{\textbf{\Large d}}}
\end{picture}
\caption{Dependence of the inference quality on the choice of the time-bin $dt$, for a purely excitatory network (same parameters as in Fig.~\ref{figure1}) with a distribution of synaptic delays $\delta{}$ between $0.1$ and $20$ ms (see Methods). Panels a-c show the distribution of inferred synaptic couplings for three values of $dt$, including $dt = 7\,\mathrm{ms} \simeq \langle \delta \rangle$; solid (dashed) lines are the distributions of inferred couplings for existing (non-existing) synapses. Panel d shows the inferred coupling $J_{Inf}$ against the associated delay $\delta$, for $dt = 7$ ms. Neurons fire at about $50$ Hz; the total recording length is $500$ s.}
\label{figure2}
\end{center}
\end{figure}
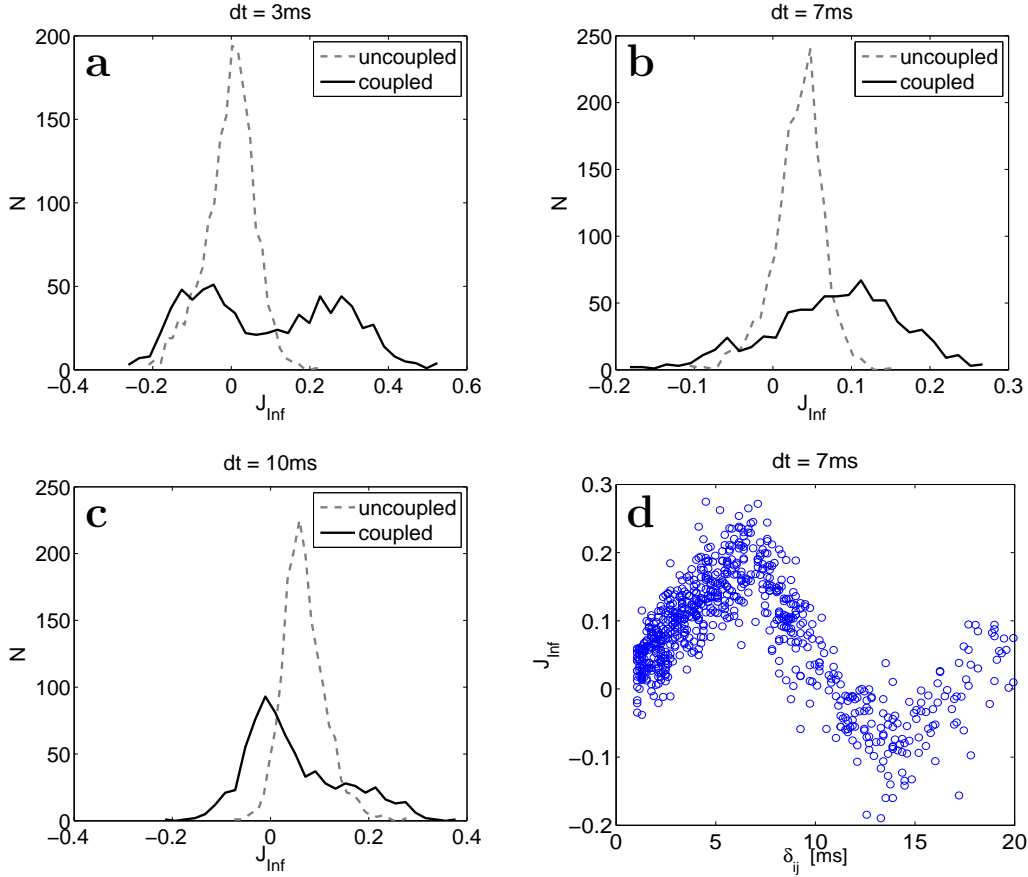

Fig.~\ref{figure1}d gives a quantitative representation of the inference quality in terms of ROC curves for the three choices of the time-bin $dt$. The fraction of correctly identified existing synapses (``true positive'') against the fraction of correctly identified unconnected pairs (``true negative'') is plotted parametrically varying a discrimination threshold: the dashed line, corresponding to $dt = 3$ ms, clearly surmounts the other two, to all effects allowing for a perfect discrimination. 
A puzzling feature can be recognized when comparing the plots in Fig.~\ref{figure1}: the histograms corresponding to existing (and always excitatory) synapses are centered around different values, depending on $dt$: in particular, for $dt < 3$ ms they appear to have been estimated as inhibitory synapses. We will come back to this point later, when explaining results in Fig.~\ref{figure2}.

We next considered the more interesting case in which delays $\delta$ were exponentially distributed between $1$ and $20$ ms (see Methods), for three values of the time-bin $dt$, including $dt = 7\,\mathrm{ms} \simeq \langle \delta \rangle$ (Fig.~\ref{figure2}a-c). The quality of the inference stays poor in all cases. Interestingly, a bimodal histogram appears for existing synapses for $dt \lesssim \langle \delta \rangle$, the two peaks being associated with $dt < \delta$ (left peak, analogous to the one seen in Fig.~\ref{figure1}a and $dt > \delta$ (right peak). To further investigate this, in Fig.~\ref{figure2}d, we plot the inferred coupling $J_{Inf,ij}$ against the associated delay $\delta$, for $dt = 7$ ms. A clear non-monotonic behavior is observed. The inferred $J_{Inf}$ is maximum when its associated delay $\delta$ equals $dt$. This can be understood observing that, for $\delta = dt$, an (excitatory) pre-synaptic spike emitted during time-bin $t$ will always reach neuron $i{}$ in the following time-bin $t + dt$, thus maximally contributing to the conditional probability $\langle S_i(t + dt) | S_j(t) = 1 \rangle$ in Eq.~(\ref{eq.jInferApprox1}), and therefore to $J_{Inf}$; for $\delta < dt$ such probability roughly scales with $\delta/dt$, assuming uniform probability of pre-synaptic spike occurrence in each time-bin $dt$ (see the initial rising ramp in Fig.~\ref{figure2}d). Analogously, for $\delta = 2\,dt = 14$ ms, the conditional probability $\langle S_i(t + 2\,dt) | S_j(t) = 1 \rangle$ will be maximum; since the probability for the post-synaptic neuron to fire in adjacent time-bins is negligible, this implies that $\langle S_i(t + dt) | S_j(t) = 1 \rangle$ will attain a minimum, thus explaining the negative peak at $\delta = 2\,dt$. The downward ramp for $dt < \delta < 2\,dt$ linearly interpolates between the two peaks, confirming the expectation that the probability of firing two time-bins after the pre-synaptic spike roughly increases as $(\delta-dt) / dt$, as the probability of firing in the previous bin decreases by the same amount. Beyond $\delta = 2\,dt$, the conditional probability $\langle S_i(t + dt) | S_j(t) = 1 \rangle$ stays below $m_i{}$ ($J_{Inf} < 0$), approaching asymptotically this value for $\delta$ greater than the average post-synaptic inter-spike interval (which is roughly $20$ ms for the data shown in Fig.~\ref{figure2}).

From the above discussion it is then clear that a positive real synaptic efficacy can result in a positive or negative inferred coupling depending on the relationship between $\delta$ and the time-bin $dt$; this explains the already mentioned negative portion of the histograms shown in Fig.~\ref{figure1}a-c. 

\subsection*{An extended Kinetic Ising Model with a distribution of synaptic delays}
The poor results obtained in Fig.~\ref{figure2} motivated us to extend the model to account for a distribution of delays $\delta_{ij}$:
\begin{equation}
H_i(t)=h_i(t)+\sum_{j}J_{ij}S_j(t -\delta_{ij})
\label{Hlocale_nuovo0}
\end{equation}
Instead of attempting to maximize the log-likelihood of the model on the data to infer $\delta_{ij}$, we devised an alternative way to estimate them from the observed neural activity, and insert them as fixed parameters in the maximum-likelihood inference of the couplings $J_{ij}$. 

\begin{figure}[h!]
\centering
\setlength{\unitlength}{\textwidth}
\begin{picture}(1,0.5)
  \put(0.12,0.)
  {
    \epsfig
    {
	file=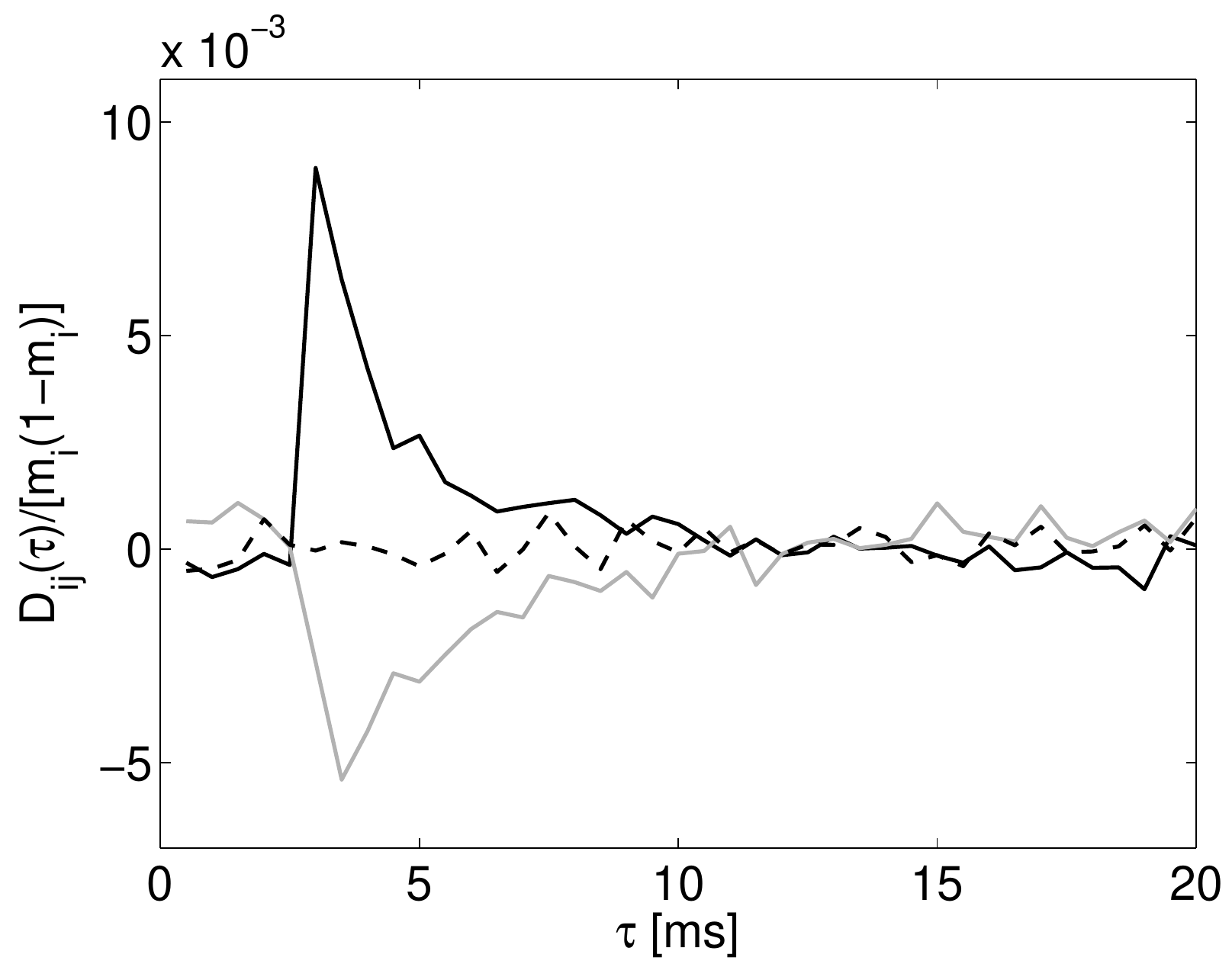,
	height=0.5\unitlength
    }
  }
\end{picture}
\caption{Dependence of cross-correlation $D_{ij}$ on $\tau$ for a fixed pair of neurons $(i,j)$. Excitatory synapse (thick solid line, synaptic delay $\delta = 3$ ms), inhibitory synapse (thin solid line, synaptic delay $\delta = 3$ ms), no synapse (dashed line), in a network with $N_E = 25$ excitatory neurons, $N_I = 25$ inhibitory neurons, connection probability $c = 0.1$, $J_{ij} = \pm 0.54$ mV, $\delta_{ij}$ exponentially distributed from $1$ ms to $20$ ms (see Methods).}
\label{figure3}
\end{figure}

The procedure is based on the intuition that the time-retarded cross-correlation, $D_{ij}(\tau)$ (see Eq.~(\ref{eq.DIjTau})), between the activities of a given pair  $ij$ of connected neurons should peak at a time-lag $\tau$ close to the actual synaptic delay $\delta_{ij}$; the peak is expected to be positive or negative depending on the synapse being excitatory or inhibitory, respectively.  This intuition is confirmed in Fig.~\ref{figure3}, in which such correlation is reported for one pair of neurons connected by an excitatory synapse (black solid line), one pair of disconnected neurons (dashed line) and one pair of neurons connected by an inhibitory synapse (solid grey line); for both the existing synapses the delay is $\delta_{ij} = 3$ ms. 

As expected, $D_{ij}(\tau)$ fluctuates around zero for disconnected neurons, showing sharp positive or negative peaks at $\tau \simeq 3$ ms, for excitatory and inhibitory transmission, slowly decaying after the peak. This latter feature is explained considering that (taking the solid black curve as an example), for $\tau > \delta_{ij}$ an excitatory spike, which was ineffective to trigger a post-synaptic spike, still depolarized the membrane, thereby increasing the firing probability on a time scale comparable with the membrane time constant. 
On the other hand, for $\tau < \delta_{ij}$, the `absence' of the contributed excitatory pre-synaptic spike, lowers (a bit, but for a time of the order of the average inter-spike interval of the pre-synaptic neuron) the firing probability of the post-synaptic neuron.  Including the analogous argument for inhibitory synapses, in summary one expects that for $\tau < \delta_{ij}$ the cross-correlation stays slightly above zero for inhibitory synapses, and slightly below zero for excitatory ones, both for a time comparable to the average inter-spike interval.

The above argument also helps explaining some highlighted features in Figs.~\ref{figure1} and \ref{figure2}. The standard procedure there used rests on the computation of the conditional probabilities $\langle S_i(t + dt) | S_j(t) = 1 \rangle$, or equivalently (see Eq.~(\ref{eq.DIjTau})) of the time-retarded correlations $D_{ij}(\tau = dt)$; with reference to the discussion at the end of the previous section, it is understood that, for a time-bin $dt$ smaller than the actual delay, the negative correlation implied by the above discussion would result in an excitatory synapse being inferred as an inhibitory one.

\begin{figure}[h!]
\centering
\setlength{\unitlength}{\textwidth}
\begin{picture}(1,0.375)
  \put(0.03,0.)
  {
    \epsfig
    {
	file=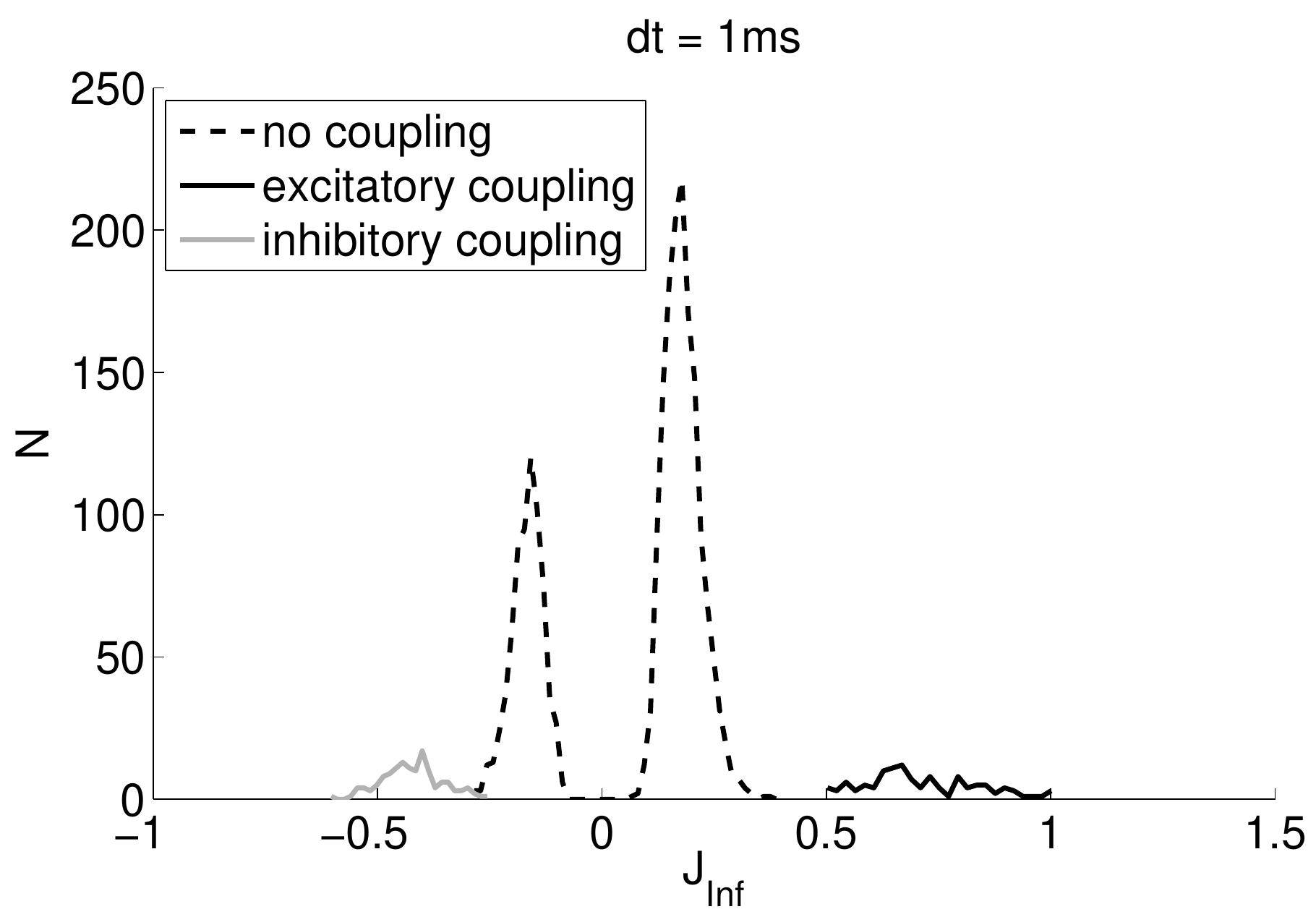,
	height=0.375\unitlength
    }
  }
  \put(0.48,0.01)
  {
    \epsfig
    {
	file=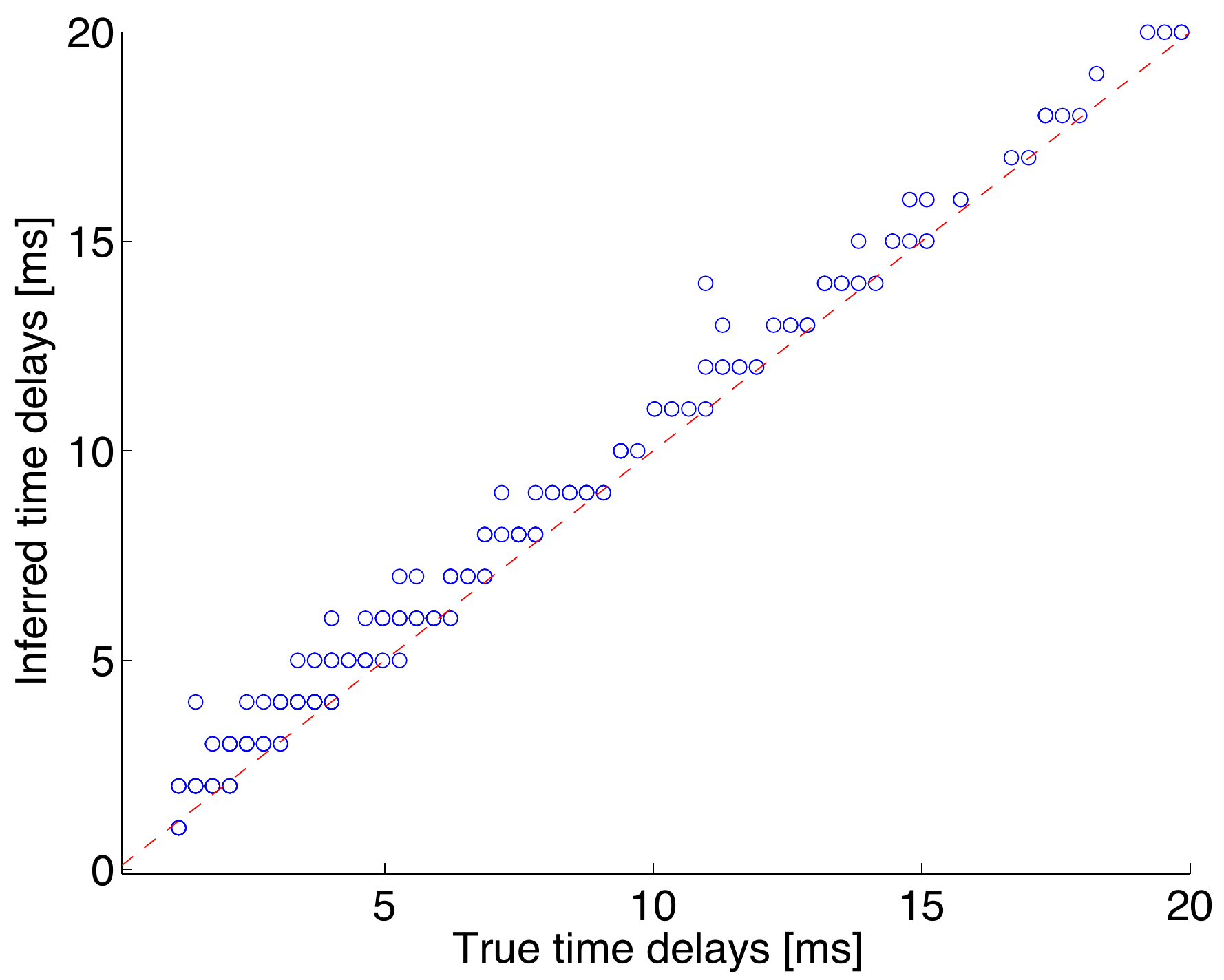,
    height=0.35\unitlength
    }
  }
\end{picture}
\caption{Inference results for the Kinetic Ising Model with time delays for a network of excitatory and inhibitory neurons (same parameters as in Fig.~\ref{figure3}), obtained with time-bin $dt = 1$ ms. Left panel: histograms of inferred synaptic couplings, for excitatory (black solid line) and inhibitory (grey solid line) synaptic contacts, and for those corresponding to unconnected pairs (dashed line). Right panel: delays inferred from the cross-correlation peaks \textit{vs} actual delays for existing synapses; the dashed line marks the identity line. Neurons fire at about $50$ Hz; the total recording length is $500$ s.}
\label{figure4}
\end{figure}

In Fig.~\ref{figure4} we show results of the above two-steps procedure for a network with both excitatory and inhibitory synapses. In the left panel, we show inferred synaptic couplings, marking with different line styles the values inferred for existing synapses (solid black for excitatory, solid grey for inhibitory ones), and disconnected neuron pairs (dashed line). Two peaks, centered around positive and negative values respectively, emerge for existing excitatory and inhibitory synapses; both peaks have null or minimal overlap with the histogram for disconnected pairs. The latter give rise to two peaks almost symmetric around zero, contrary to what observed in Figs.~\ref{figure1} and \ref{figure2}. This feature derives from the procedure for inferring the synaptic delays, where we choose the lag $\tau$ corresponding to the largest absolute value of the time-retarded cross-correlation $D_{ij}(\tau)$; such choice leads to the inference of statistically non-null synaptic couplings, even with the noisy cross-correlation found for unconnected neurons. However, the two peaks related to unconnected neuron pairs are expected to shift towards zero for longer recordings, according to the scaling appropriate for extreme value statistics.

Fig.~\ref{figure4}, right panel, shows the delays inferred from the cross-correlation peaks \textit{vs} actual delays for existing synapses; a very good match is appreciable over the extended range covered by the exponential distribution ($1-20$ ms). It is worth noticing that inferred time delays are never smaller than true ones, while they can be slightly larger due to the noise in the data and their inherent discretization as multiples of the time-bin $dt$.

\subsection*{Relation between true and estimated synaptic efficacy, and its dependence on the time-bin}
We notice from Fig.~\ref{figure4} that, although the inference procedure successfully identifies excitatory and inhibitory synapses, the corresponding distributions are centered around values of different module, while the excitatory and inhibitory synaptic efficacies were chosen in the simulation to have equal absolute values. It would be of course interesting to give a theoretical account of such asymmetry, which would also allow one to remap the inferred values onto quantitatively reliable estimates of the real synaptic efficacies.

For excitatory synapses, an intuitive argument can hint at a strategy of computation. Starting from Eq.~(\ref{eq.jInferApprox1}), the probability $\langle S_i | S_j = 1\rangle$ that neuron $i{}$ will fire at time $t$ ($S_i(t) = 1$) upon receiving a spike from neuron $j$ ($S_j(t - \delta_{ij}) = 1$) is roughly equal to the probability that the membrane potential $V(t)$ of the post-synaptic neuron is at a distance less than the synaptic efficacy $J_{ij} > 0$ from the firing threshold $\theta$; such probability is the integral, between $\theta - J_{ij}$ and $\theta$, of the probability density $p(V)$ of the membrane potential; assuming that the rest of the network, and possibly external sources, provide a noisy input (of infinitesimal variance $\sigma^2$) such that the diffusion approximation holds \cite{renart2004mean}, the threshold is an absorbing barrier for the stochastic process $V_i(t)$, and this implies that $p(\theta) = 0$. Therefore, in stationary conditions, expanding $p(V)$ close to $\theta$ (indeed, consistently with the diffusion approximation, $J_{ij} \ll \theta - V_{rest}$, being $V_{rest}$ the equilibrium value of the membrane potential absent any external input):

\begin{eqnarray}
\langle S_i | S_j = 1\rangle & \simeq & \int_{\theta-J_{ij}}^{\theta} p(v) \, dv \simeq \int_{\theta-J_{ij}}^{\theta}  p'(\theta) \, (v-\theta) \, dv = \nonumber \\
&&-\frac{2 \nu}{\sigma^{2}} \int_{\theta-J_{ij}}^{\theta}   (v-\theta) \, dv =\frac{\nu J_{ij}^{2}}{\sigma^{2}}
\end{eqnarray}
where we have used, for the stationary average firing rate, the equivalence $\nu=-(\sigma^{2}/2) p'(\theta)$. Inserting this result into Eq.~(\ref{eq.jInferApprox1}), then we have $J_{Inf} \sim J_{True}^{2} / dt$; thus (for $J_{ij} > 0$) the relationship is quadratic and divergent with decreasing time-bin $dt$.

In Methods this rough derivation is refined to take into account afferent external spikes to neuron $i$ in a single time-bin $dt$, that can make the neuron fire for $V < \theta - J_{ij}$, and even when $J_{ij} < 0$ (inhibitory synapse); due to the fact that for the latter case neuron $i$ will fire only when external spikes overcompensate the inhibitory synaptic event, the found scaling (Eq.~(\ref{Jneg1})) is different: still quadratic in $J_{ij}$, but constant in $dt$ We remark that the found relationship for inhibitory synaptic couplings basically gives $J_{Inf} \gtrsim -1/2$ whenever $J_{ij} < -J_{ext}$ where $J_{ext}$ is the synaptic efficacy of synapses from external neurons (see Eq.~(\ref{eq.jInferInhGreaterJExt})); thus the inference method loses sensitivity for $|J_{ij}|$ approaching $J_{ext}$; moreover, since the $J_{Inf}$ are estimated from the simulated data and one can actually have $J_{Inf} < -1/2$ because of noise in the estimate, no values $J_{True}$ can be inferred in these cases.

\begin{figure}[!h]
\begin{center}
\setlength{\unitlength}{\textwidth}
\begin{picture}(1,0.8)
  \put(0.02,0.4)
  {
    \epsfig
    {
    file=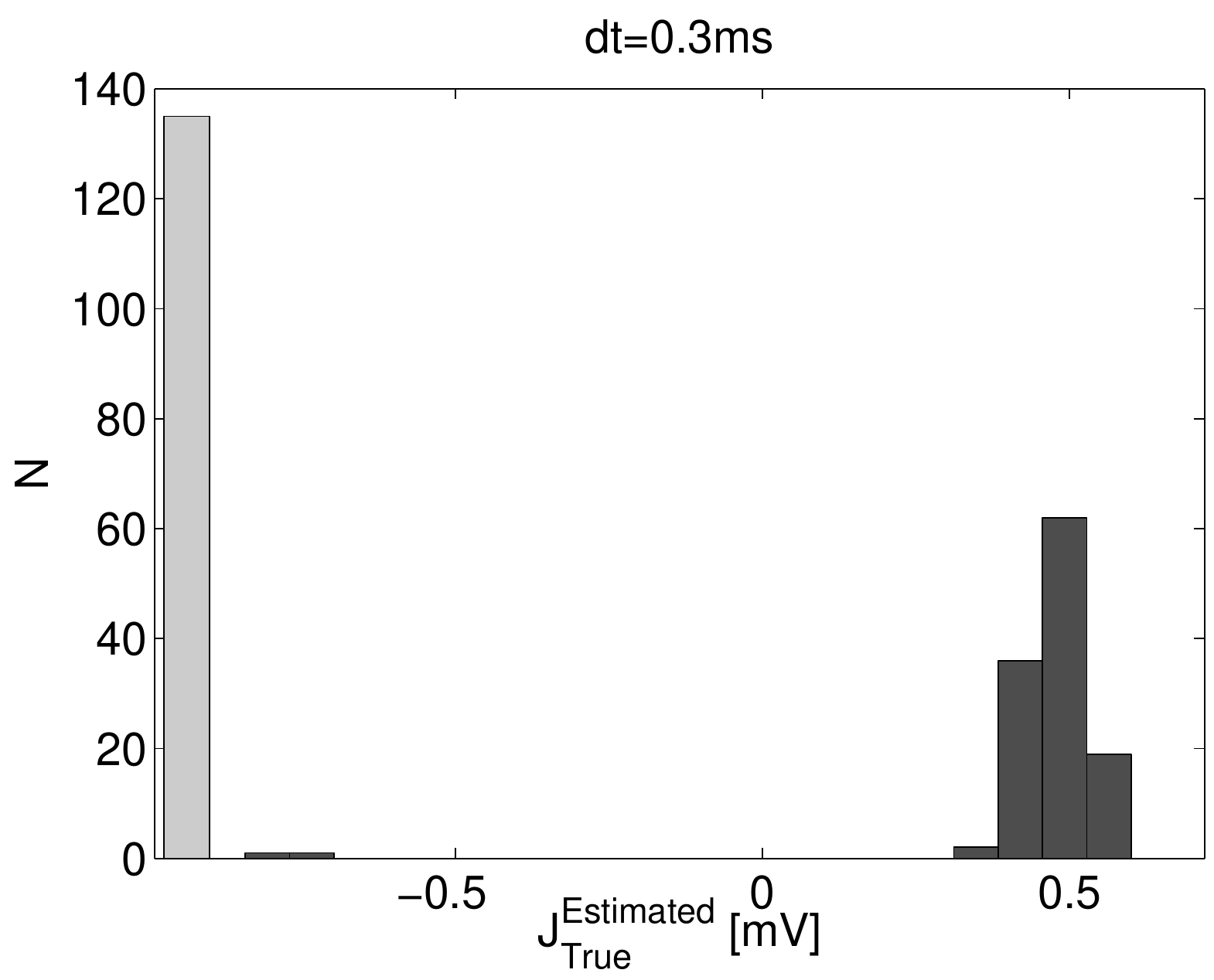,
    width=0.45\unitlength,
    }
  }
  \put(0.48,0.4)
  {
    \epsfig
    {
    file=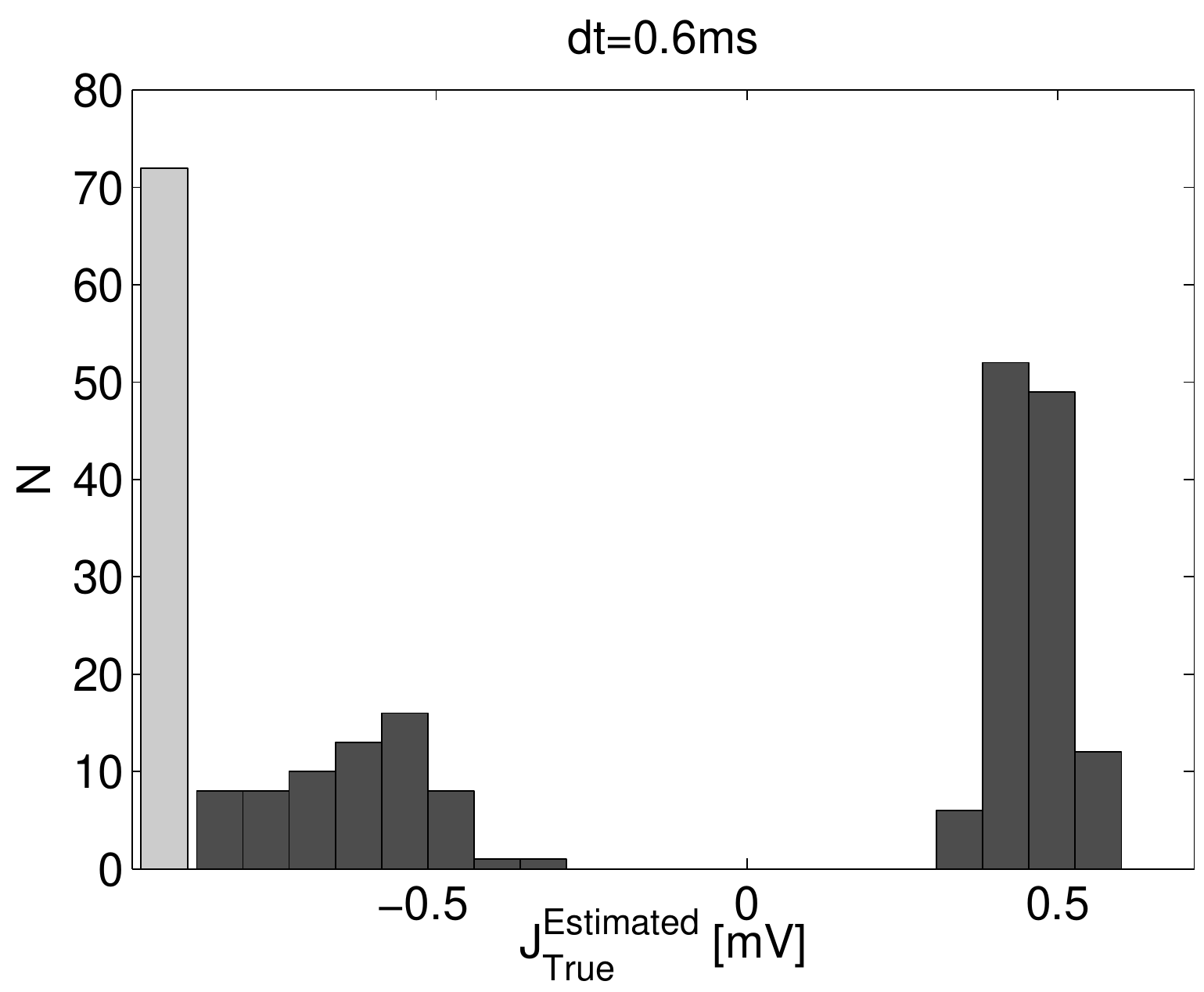,
    width=0.45\unitlength,
    }
  }
  
  \put(0.02,0.)
  {
    \epsfig 
    {
    file=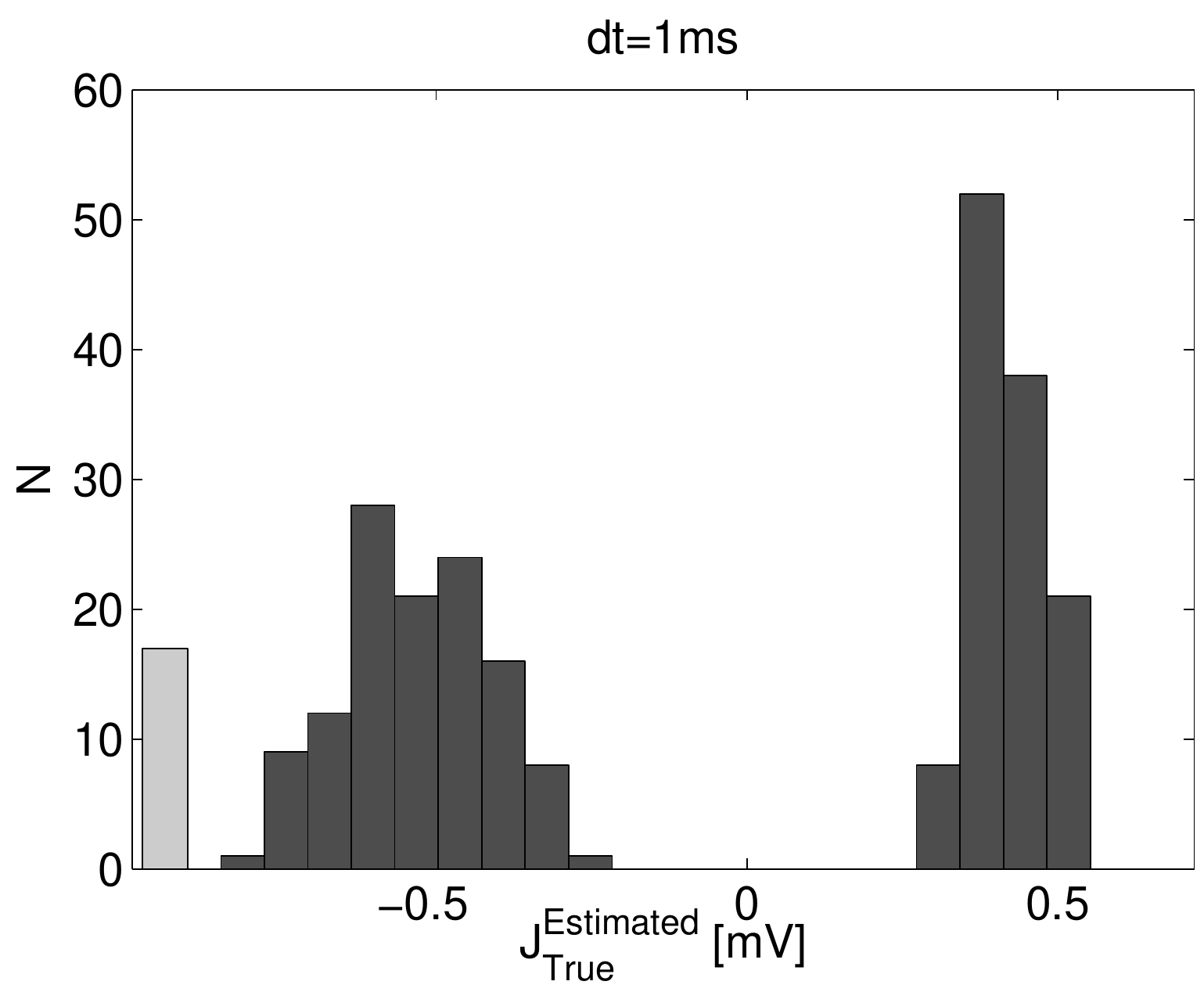,
    width=0.45\unitlength,
    }
  }
  \put(0.48,0.)
  {
    \epsfig
    {
    file=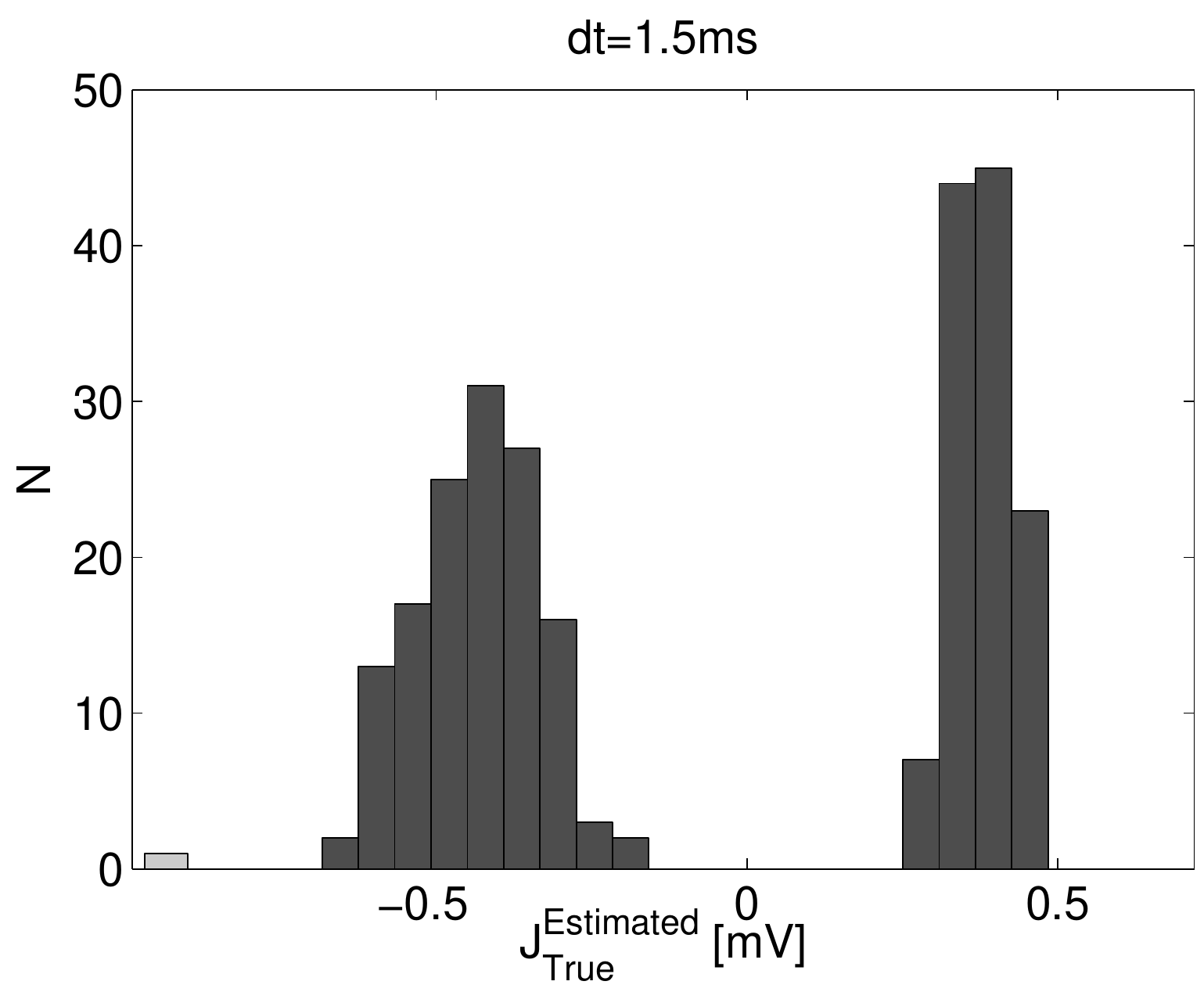,
    width=0.45\unitlength,
    }
  }
\end{picture}
\caption{Histograms of $J_{True}^{Estimated}$: theoretical estimates of $J_{True}$ from the inferred coupling $J_{Inf}$ for the same network of Fig.~\ref{figure4}. The four panels show the histograms of $J_{True}^{Estimated}$ (darker bars) for existing synapses only, for different choices of the time-bin $dt$; for excitatory and inhibitory synapses Eq.~(\ref{Jpos1}) and Eq.~(\ref{Jneg1}) are applied, respectively. The two peaks already seen in Fig.~\ref{figure4}, corresponding to excitatory and inhibitory neurons, are now expected to center around the true values $J_{True} = \pm 0.54$ mV. The light-grey bar in each histogram represents the number of inhibitory synapses for which the found value of $J_{Inf}$ fell below the lower bound of the physical range, $-\frac{1}{2} < J_{Inf} < 0$, outside of which the relationship in Eq.~(\ref{Jneg1}) has no inverse solution; the bar is arbitrarily placed just on the left of the minimum value for $J_{True}^{Estimated}$ obtainable by inverting Eq.~(\ref{Jneg1}).}
\label{figure5}
\end{center}
\end{figure}

In the following, we call $J_{True}^{Estimated}$ the value obtained by inverting the relationship between $J_{Inf}$ and $J_{True}$.
Fig.~\ref{figure5} shows the histograms of $J_{True}^{Estimated}$ for four values of the time-bin $dt$, for the same network of Fig.~\ref{figure4}. Since $J_{True} = \pm 0.54$ mV in this network, the expectation is to find two peaks, one for excitatory and one for inhibitory synapses, around these two values. This expectation is substantially confirmed by the shown results for excitatory synapses, with a slight gradual worsening for increasing time-bin $dt$; such worsening can be understood by noting that the relationship Eq.~(\ref{Jpos1}), valid for excitatory synapses, is derived for $dt \rightarrow 0$. 
We notice that, for instantaneous synaptic transmission, for $J_{ij} > 0$ every incoming pre-synaptic spike from neuron $j$ has a probability of making neuron $i{}$ fire that does not vanish as $dt \rightarrow 0$; this is not true for inhibitory synapses, for which what matters is the number of Poisson excitatory events in the time-bin $dt$ (from other neurons) needed to overcompensate the effect of an inhibitory pre-synaptic spike in the same time-bin $dt$. The probability of having this number in $dt$ vanishes with $dt$, thereby making the number of observed post-synaptic spikes following a pre-synaptic spike vanish with $dt$; the correlation-based estimates $J_{Inf}$ are therefore affected by greater noise. The noise broadens the distribution of $J_{Inf}$ and populates a tail of un-physical values $J_{Inf} < -\frac{1}{2}$, as discussed above, which contributes to the grey bar in the shown histograms.

The above considerations highlight the existence of a trade off in the choice of the time-bin $dt$, if one wants to derive a reliable estimate of true excitatory and inhibitory synaptic values from the ones inferred through the Kinetic Ising Model.
A too small time-bin $dt$ does not allow to infer inhibitory synapses, while a too large $dt$ introduces a systematic bias in the estimated synaptic efficacies.

\begin{figure}[h!]
\centering
\setlength{\unitlength}{\textwidth}
\begin{picture}(1,0.375)
  \put(0.04,0.)
  {
    \epsfig
    {
	file=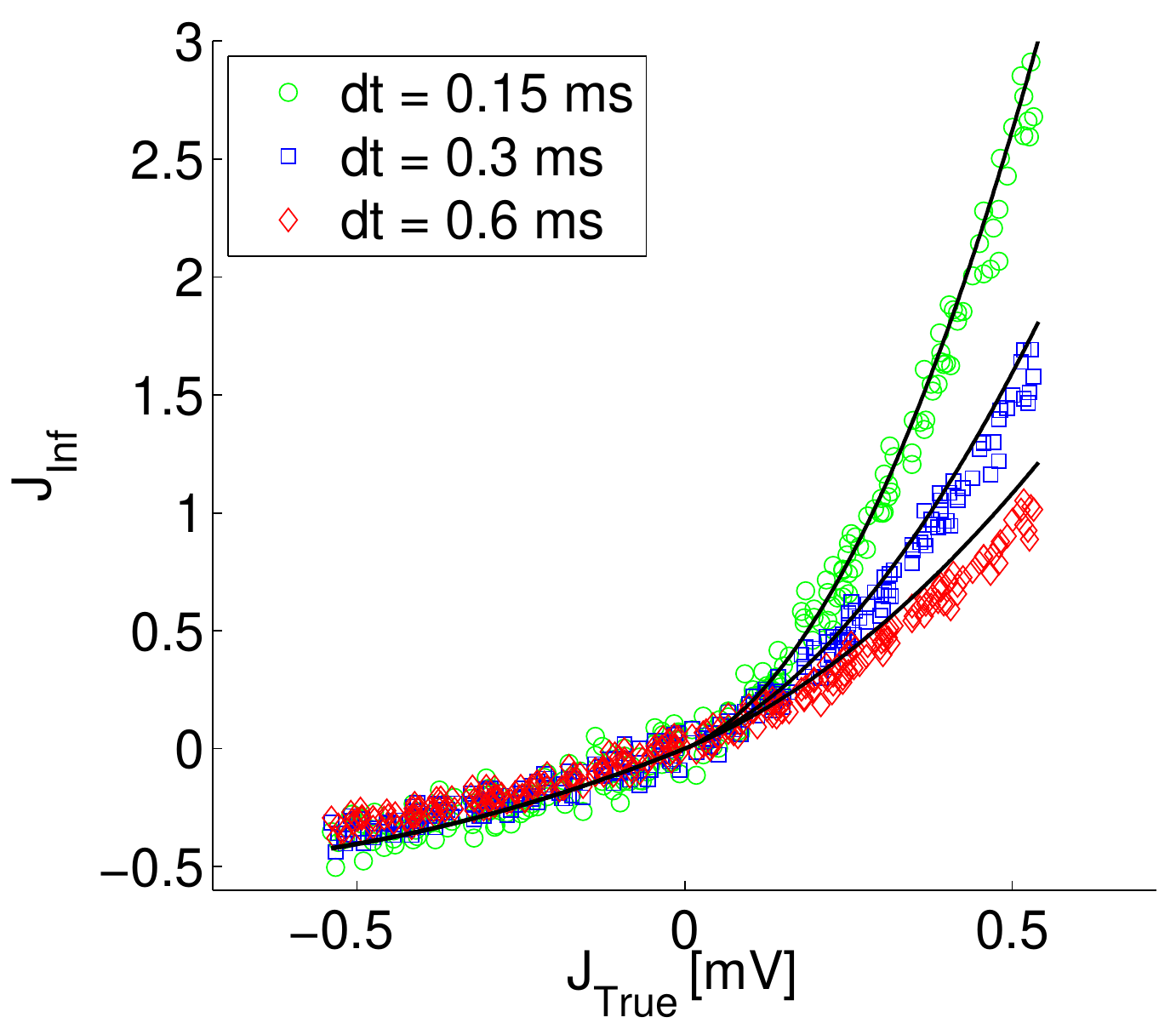,
	height=0.375\unitlength
    }
  }
  \put(0.49,0.005)
  {
    \epsfig
    {
	file=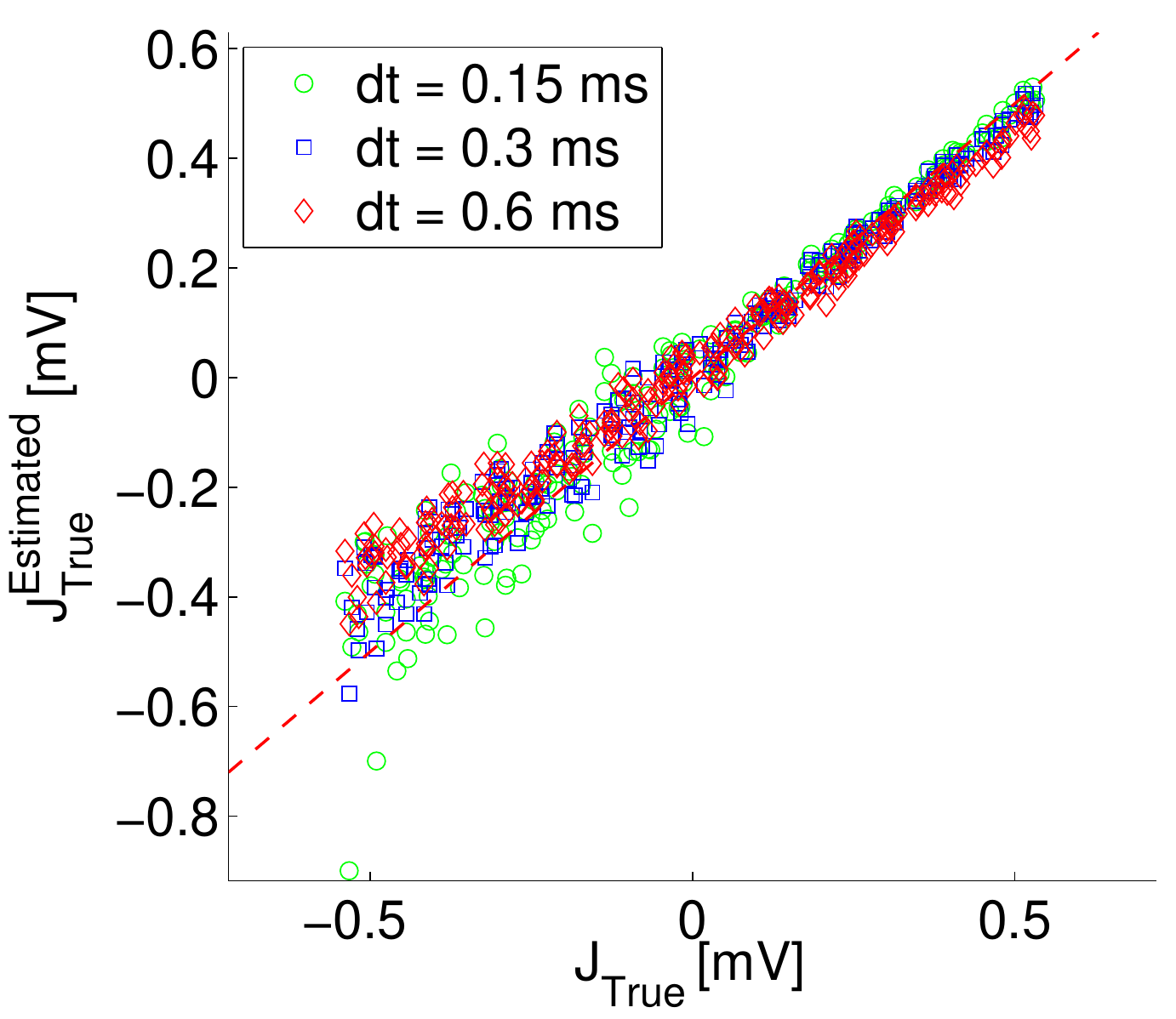,
    height=0.35\unitlength
    }
  }
\end{picture}
\caption{Relationship between inferred couplings and synaptic efficacies for varying time-bin $dt$. Left panel: $J_{Inf}$ \textit{vs} $J_{True}$. Right panel: $J_{True}^{Estimated}$ \textit{vs} $J_{True}$. In both cases only non-zero synapses were included. Results are for a network of $N_E = 25$ excitatory neurons and $N_I = 25$ inhibitory neurons, sparsely connected with probability $c = 0.1$, and firing at about $20$ Hz; the synaptic efficacies are uniformly distributed in $-0.54$ mV $ < J_{ij} < 0.54 $ mV; $\delta_{ij} =3$ ms.}
\label{figure6}
\end{figure}

To illustrate the predicted scaling of $J_{Inf}$ \textit{vs} $J$ with respect to the time-bin $dt$, we performed simulations with a uniform distribution of synaptic efficacies and three values of $dt$. In Fig.~\ref{figure6}, left panel, we plot the inferred synaptic couplings against the real synaptic efficacies. All the predicted features are nicely matched: the quadratic dependence $J_{Inf} \sim J^2$, the strong dependence on $dt$ for excitatory synapses (divergence for $dt \rightarrow 0$), and the approximate independence of the inhibitory couplings from $dt$. The above theoretical predictions are used to rescale the inferred synaptic couplings, which are compared to the true ones in the right panel of Fig.~\ref{figure6}. The validity of the approximation is confirmed by the collapse of the three curves on the main diagonal.

\subsection*{Inference on a bursting network with spatial structure}
So far, the simulated networks were uniformly sparsely connected, with no spatial structure; moreover, the neural activity was stationary and asynchronous. As a step towards checking the robustness of the method in more complex situations, we simulated a network with the following spatial structure. The $N = 1000$ excitatory neurons are subdivided into $P = 100$ populations, organized on a circle; the probability $c_{\alpha\beta}$ ($\alpha,\,\beta = 1,\, \dots,\, P$) that a neuron in population $\beta$ is pre-synaptic to a neuron in population $\alpha$ is $c_{\alpha\beta} = 0.134 \; e^{-\frac{d(\alpha,\,\beta)}{135} - \frac{d(\alpha,\, P / 2)}{26.1}}$, where $d(\alpha,\,\beta) = \min(|\alpha - \beta|,\,P - |\alpha - \beta|)$ is the distance along the circle. Therefore, each population is more connected to its immediate neighbors, and some populations receive more pre-synaptic synapses than others; the average connection probability is $5\%$. Besides, the excitatory synaptic efficacies have Gaussian distribution.

\begin{figure}
\centering
\setlength{\unitlength}{\textwidth}
\begin{picture}(1,0.75)
  \put(0.075,0.375)
  {
    \epsfig
    {
	file=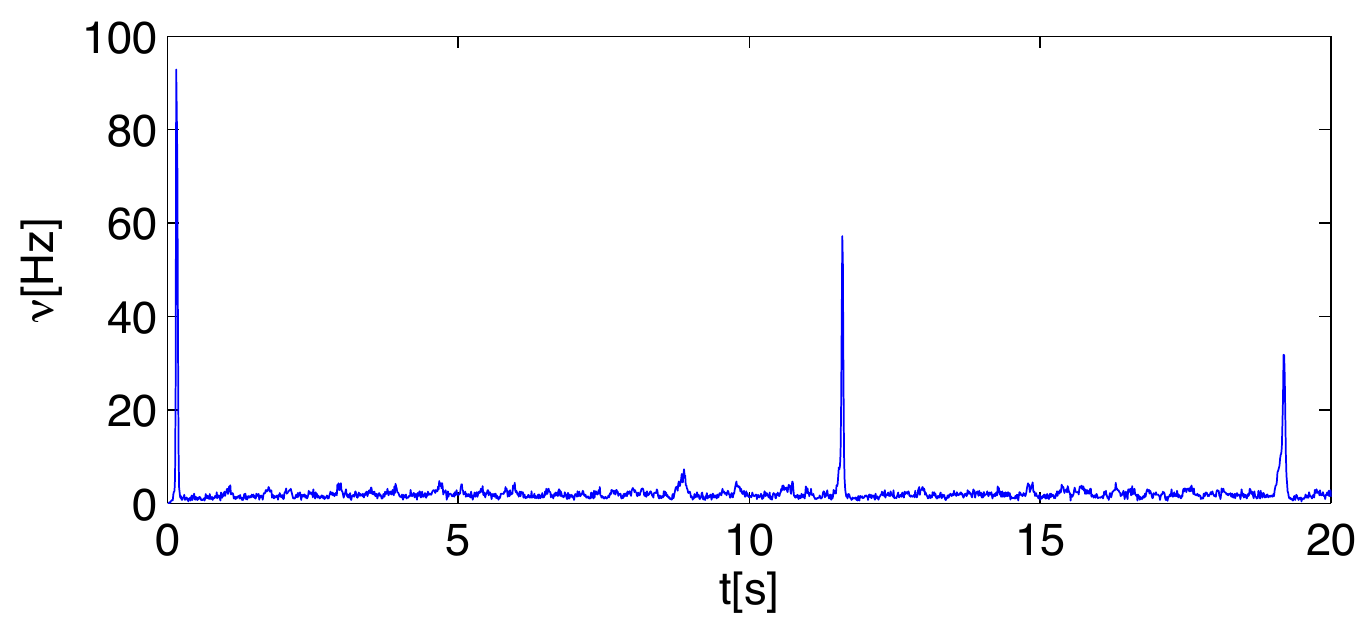,
    width=0.8\unitlength
    }
  }
  \put(0.075,0.0)
  {
    \epsfig
    {
	file=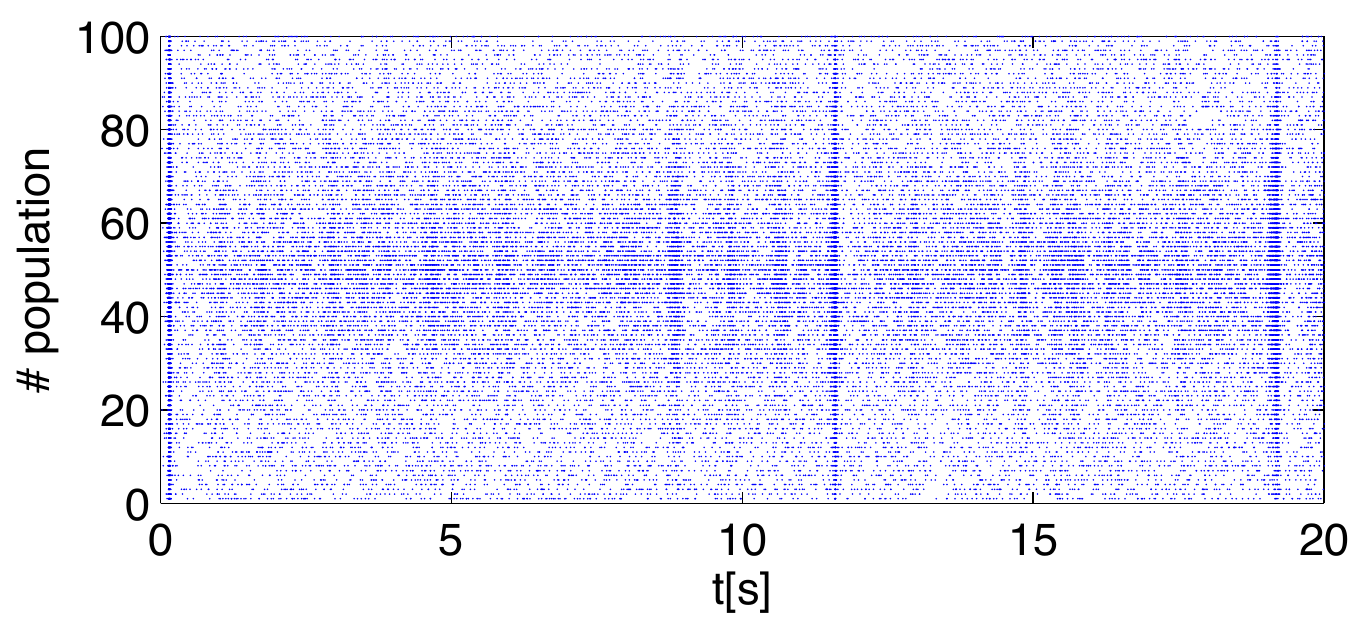,
    width=0.8\unitlength
    }
  }
\end{picture}
\caption{Illustrative time course of neuronal activity in a bursting network with spatially non-uniform synaptic connectivity and synaptic short-term depression (see text for details). Top panel: sample time course of the average firing rate. Bottom panel: raster plot from a sub-set of 100 neurons, corresponding to the time record of the top panel. The network comprises $N = 1000$ excitatory neurons ($J_{ij}$ are drawn from a Gaussian distribution with mean $0.846$ mV and standard deviation of $0.211$ mV), $\delta_{ij} $ exponentially distributed from $1$ ms to $15$ ms, and synapses are endowed with short-term depression with recovery time $\tau_r = 800$ ms and synaptic release fraction $u = 0.2$ (see Methods for details); the external current is a train of Poisson spikes with average rate $\nu_{ext} = 0.6$ kHz, and synaptic efficacy randomly sampled from a Gaussian distribution of mean $J_{ext} = 0.9$ mV and standard deviation $0.225$ mV.}
\label{figure7}
\end{figure}

The neurons are furthermore endowed with short-term synaptic depression, implemented according to the model in \cite{tsodyks1997neural}. Such mechanism, mediating a self-inhibiting effect on the collective firing of the network, can generate short-lived bursts of activity, followed by periods of quiescence \cite{holcman2006emergence}, in response to random fluctuations of the overall activity, analogously to what is often observed in neuronal cultures \cite{eytan2006dynamics}. We emphasize that the network capable of spontaneously generated bursts is close to the instability of the low-activity asynchronous state, thus making the activity, even when restricted to the inter-burst periods, more correlated than the one of networks used in the preceding sections: for this more excitable network, correlated fluctuations (even in the low state) constitute a kind of global component of the activity that makes the contribution of the single synaptic contact harder to detect in the correlation functions. Fig.~\ref{figure7} shows an illustrative time course of the non-stationary, bursting neural activity from simulations.

\begin{figure}[h!]
\begin{center}
\setlength{\unitlength}{\textwidth}
\begin{picture}(1,0.35)
  \put(0.49,0)
  {
    \epsfig
    {
    file=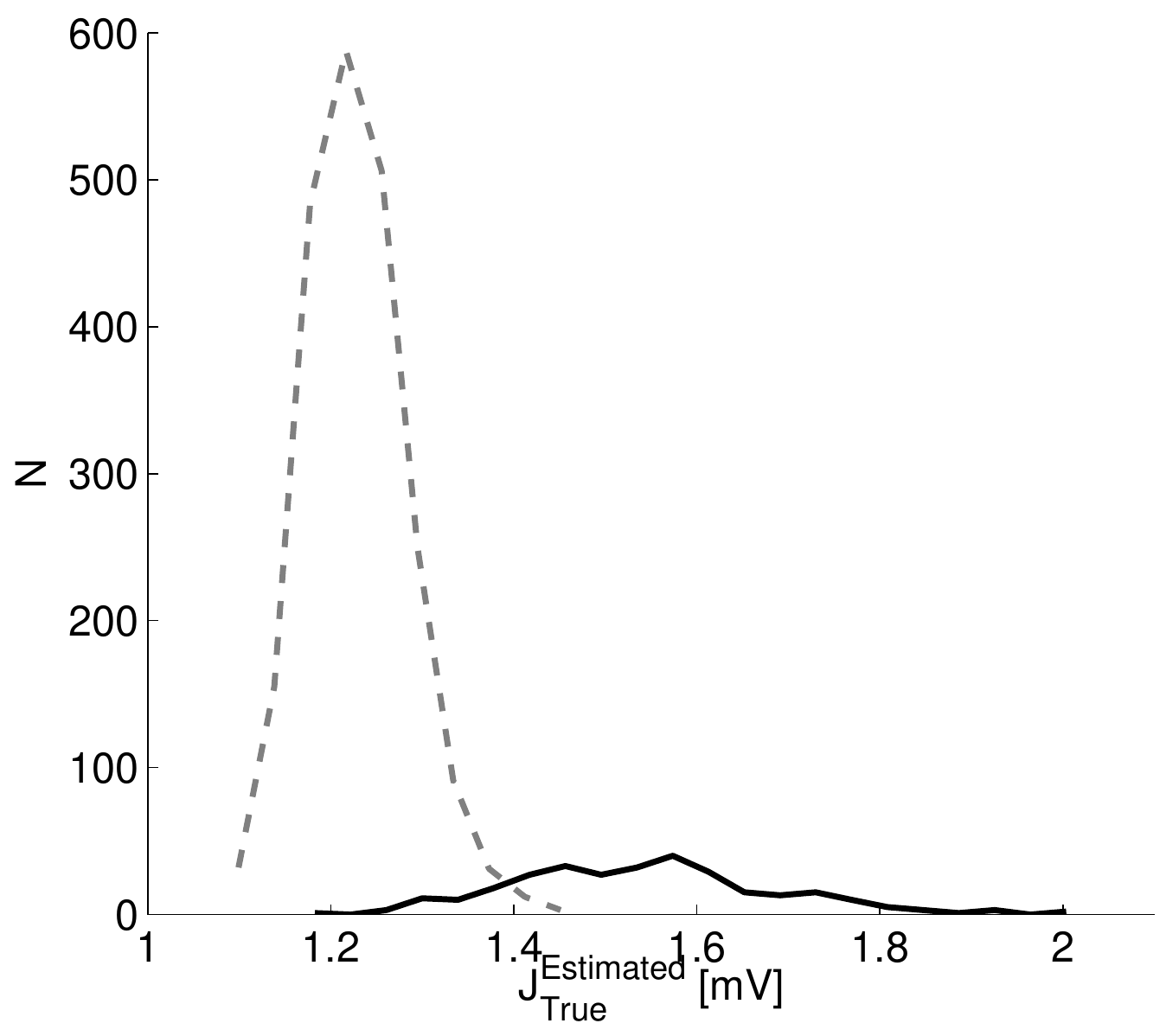,
    width=0.45\unitlength,
    }
  }
  \put(0.01,0)
  {
    \epsfig
    {
    file=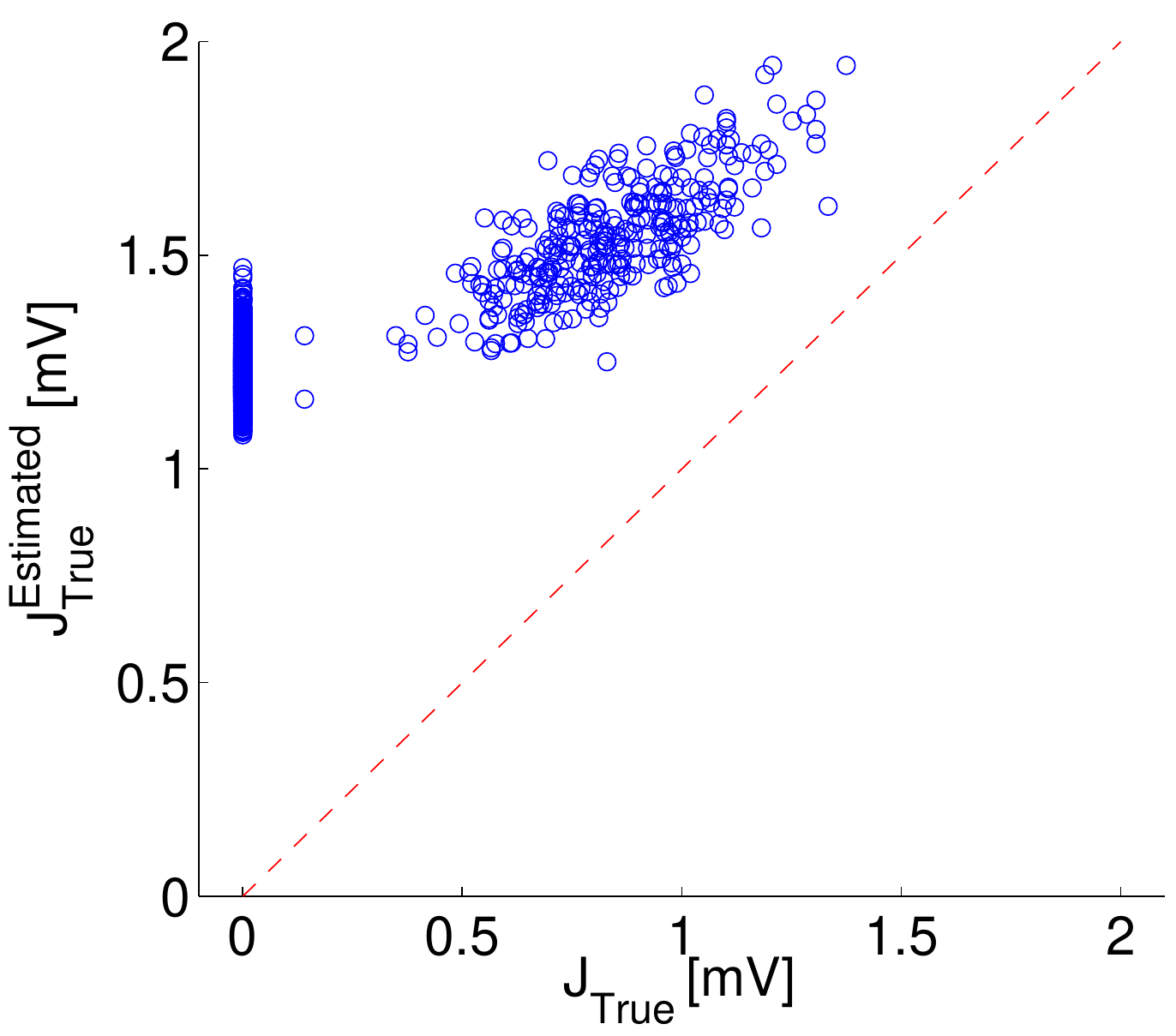,
    width=0.45\unitlength,
    }
  }
\end{picture}
\caption{Inference results for the Kinetic Ising Model with time delays for the network described in Fig.~\ref{figure7}. The inference is carried out on the 50 (out of 1000) most active neurons in the network, with $dt = 1$ ms, over a recording time of about 11 hours. Delays inferred from the cross-correlation peaks \textit{vs} actual delays (not shown) for existing synapses have $R^2 = 0.976$ with the identity line. Left panel: scatter plot of $J_{True}^{Estimated}$ \textit{vs} $J_{True}$, where $J_{True}^{Estimated}$ is computed using Eq.~(\ref{Jpos1}) (purely excitatory network) and then rescaled $J_{ij}^{Estimated} \rightarrow J_{ij}^{Estimated} \, (1 + u\,\tau_r\,\nu_j)$, where $\nu_j$ is the average firing rate of the pre-synaptic neuron, in order to compensate for the average synaptic depression induced by short-term depression (see Eq.~(\ref{eq.rSTDStationary})); the dashed line marks the identity line.  Right panel: histograms of $J_{True}^{Estimated}$ for existing synapses (solid line) and unconnected pairs (dashed lines); we remind that the $J_{True}$ have Gaussian distribution with $\langle J_{True} \rangle = 0.846$ mV.}
\label{figure8}
\end{center}
\end{figure}

In Fig.~\ref{figure8} we show the results of the inference procedure carried out on the subset of the $50$ most active neurons in the network; the time record used for inference is of about $11$ hours, with $dt = 1$ ms.

We note that for stationary pre-synaptic activity $\nu_j{}$, synaptic short-term depression effectively reduces $J_{ij}$ to $J_{ij} \langle r_{ij} \rangle = J_{ij} / (1 + u\,\tau_r\,\nu_j)$ (see Eq.~(\ref{eq.rSTDStationary})) where $r_{ij}$ is the fraction of synaptic resources available and $\tau_r$ is its recovery characteristic time scale. We corrected $J_{True}^{Estimated}$ by this factor.

The obtained inference is good for synaptic delays (not shown), while $J_{True}^{Estimated}$ are clearly off target, with a large overestimation of synaptic efficacies. Besides, the right panel of Fig.~\ref{figure8} shows a large overlap between the $J_{True}^{Estimated}$ for existing and non-existing synapses.
We remark that, with respect to the hypothesis underlying the approximate, static mean-field formulation we adopted, the bursting regimes violates all of them: correlated fluctuations, non-stationarities, and large deviations from the average firing rates. We notice also that all coherent dynamic components contribute to overestimate the synaptic couplings.

\begin{figure}[h!]
\begin{center}
\setlength{\unitlength}{\textwidth}
\begin{picture}(1,0.35)
  \put(0.49,0)
  {
    \epsfig
    {
    file=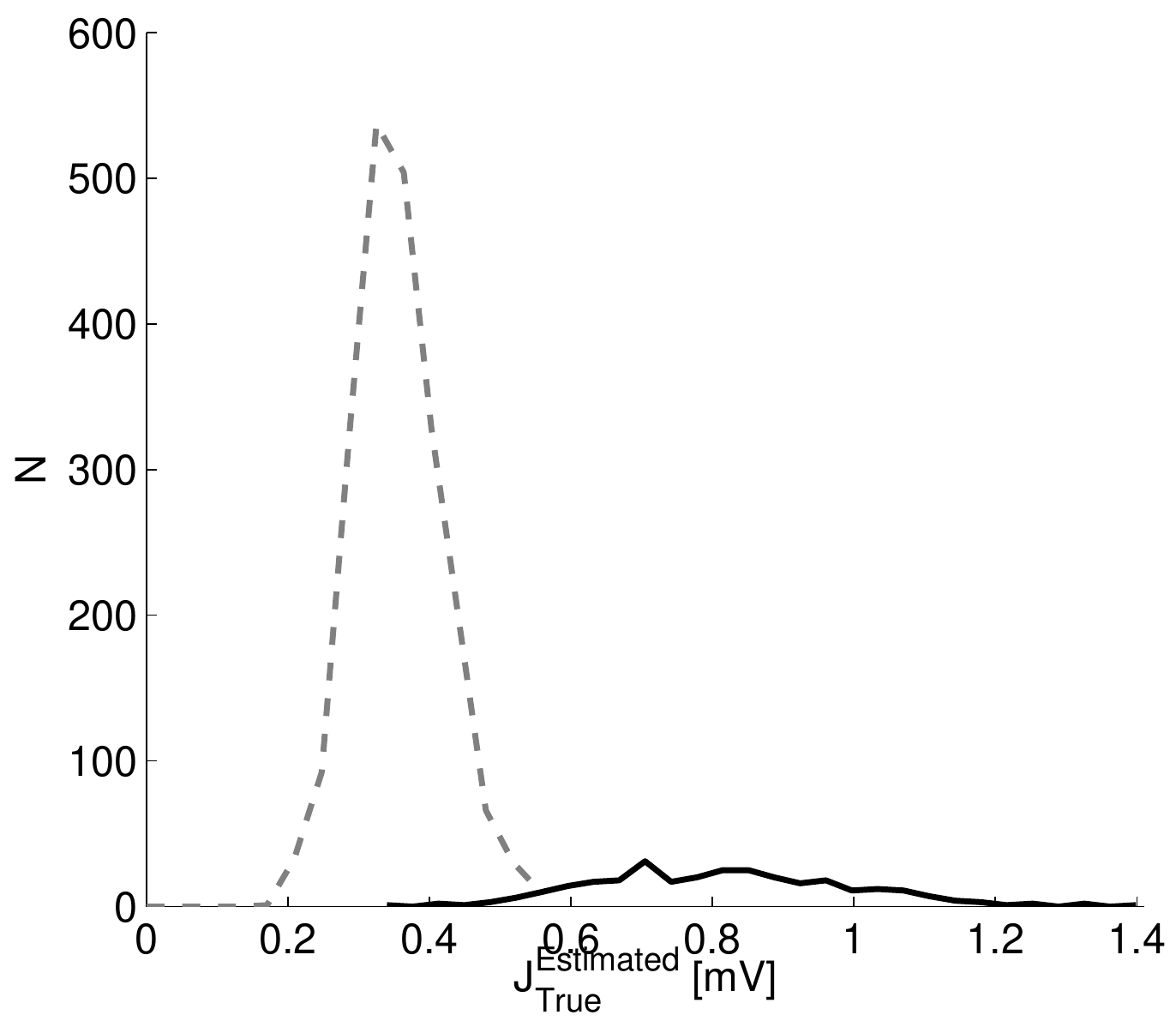,
    width=0.45\unitlength,
    }
  }
  \put(0.01,0)
  {
    \epsfig
    {
    file=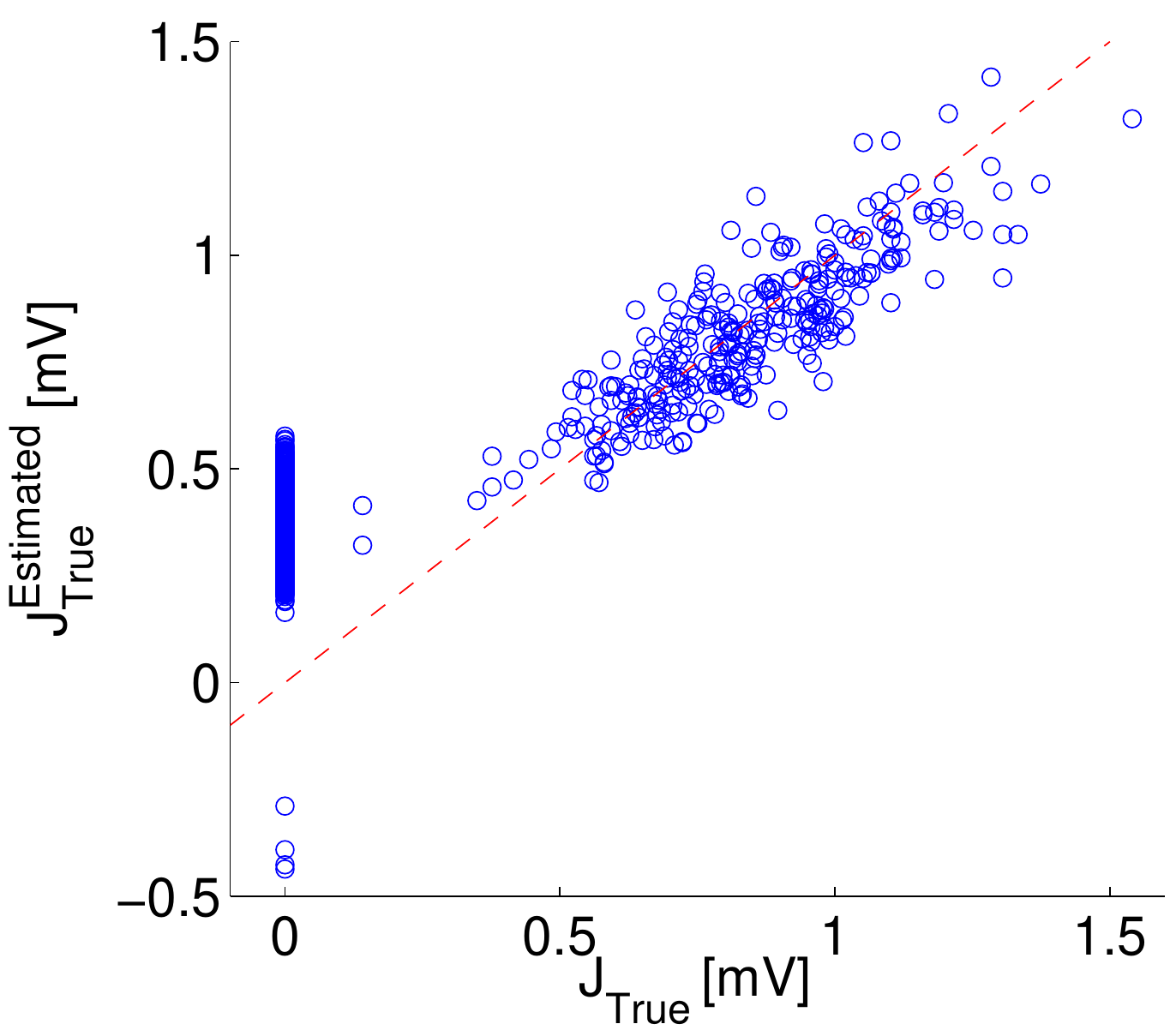,
    width=0.45\unitlength,
    }
  }
\end{picture}
\caption{Inference results for the Kinetic Ising Model with time delays for the network described in Fig.~\ref{figure7}; all the details are as in Fig.~\ref{figure8}, with the exception that the inference procedure is carried out considering only the inter-burst (low activity) periods during a recording time of 2 hours only (about $1/5$ of the recording length used in Fig.~\ref{figure8}). Delays inferred from the cross-correlation peaks \textit{vs} actual delays for existing synapses have $R^2 = 0.944$ with the identity line.}
\label{figure9}
\end{center}
\end{figure}

We therefore performed an inference restricted to the inter-burst periods, considering this time $2$ hours of simulation only. Such restriction is shown in Fig.~\ref{figure9} to greatly improve the quality of the inference.
We remark that, despite the restriction to the low activity component of the dynamics, performing inference on this network remains a non-trivial test with respect to the results reported in previous sections. Indeed, we have spatial structure in the synaptic connectivity, and the network expresses its high self-excitability also in the statistics of the fluctuations in the low activity state.

\section*{Discussion}

The main focus of our work was to extend existing methods for inferring synaptic couplings, based on the Kinetic Ising Model, in order to incorporate a distribution of interaction time scales in the neural network dynamics, in their relationship with the time-bin $dt$ used to discretize the data. In doing this, we derived analytically the relationship between the inferred couplings of the Ising model and the true synaptic efficacies of the spiking neural network generating the data.

The impact of the choice of the time-bin on the quality of the inference has been considered in the past. In the context of maximum entropy estimates in equilibrium models, the authors of \cite{roudi2009pairwise} argued that the needed constraints on $dt$ doom the method to poor performance for large networks.
Also in \cite{roudi2009statistical}, extending an analysis performed in \cite{roudi2009}, the authors studied the influence of the network size and time-bin on the quality of the inference, comparing the independent pair and naive mean field approximations with the results of the computationally demanding maximization of the likelihood through Boltzmann learning; the reported result was a decrease of the quality when increasing either the network size or the time-bin. While in the cited papers the model neurons used in simulations did have a characteristic time scale (associated with the kernel of the conductance dynamics) the analysis of the choice of the time-bin in relation with such a time scale, which we addressed here, was outside their scope.

We first checked the results of an established inference method based on the Kinetic Ising Model, showing that acceptable results are obtained only if the time-bin $dt$ matches the time scale for the interaction between neurons (spike transmission delay in our case).
Then we have shown that, for the more realistic case of a distribution of time scales across the network, any choices of the time-bin provide poor results (to the point that, for example, excitatory synapses can be mis-inferred as inhibitory ones).

Motivated by the above observations, we devised a two-step method to first estimate, for each sampled neurons pair, a characteristic time scale of their interaction (spike transmission delay), and then to use such estimated time scales as pair-dependent time-lags in the modified Kinetic Ising Model we introduce. The method can cope with wide distributions of time scales, these latter are reliably estimated, and synaptic couplings are inferred with good quality. 

Even if the newly proposed inference method is made largely independent from the choice of the time-bin $dt$, the numerical values of the resulting inferred couplings still depend on $dt$, with noticeable differences between excitatory and inhibitory synapses.
To resolve this issue, we studied in detail the stochastic equation controlling the evolution of the neuron's membrane potential, thus deriving an analytical expression relating inferred couplings to true synaptic efficacies.
Such a relation turns out to be always quadratic, $J_{Inf} \sim J_{{True}}^2$, and does depend critically on $dt$ for excitatory synapses. This allowed us to rescale the couplings $J_{Inf}$ to obtain a quantitatively good match with the true synaptic efficacies $J_{{True}}$.

The analytical relations we derived hold for the specific neuronal-synaptic model we considered, that is the leaky integrate-and-fire neuron with instantaneous synaptic transmission. Nonetheless, as long as the considered time-bin $dt$ is long w.r.t. the synaptic transmission times, and the membrane potential dynamics in the proximity of the spike emission is well approximated by an integrator with a firing threshold, then we do expect those expressions to give reasonable results. Such conditions are probably not too unrealistic for biological neurons and the fast components of the synaptic transmission. However, we remark that for real neural data a quantitative assessment of the relationship between inferred connections and synaptic strengths is still not resolved, and will probably require an incremental refinement of the inference procedure based on increasingly realistic models.

An attempt was made in \cite{cocco2009neuronal} to infer from spikes data the synaptic efficacies of non-leaky integrate-and-fire neurons (not the couplings of an Ising model);  this was done by directly maximizing the likelihood of the paths travelled by the membrane potential of a post-synaptic neuron between subsequent spikes it emits. For the sake of analytical tractability approximations were adopted (to our understanding, the most consequential being the absence of a leakage term in the neuron dynamics).
Though we did not perform a systematic comparison between the two methods, which depends on several factors (possibly including the more or less noisy firing regime of the network), we did check in some cases of interest, including the one showed in Fig. 6,  and the method proposed in \cite{cocco2009neuronal} provides worse performance (in particular, the distribution of inferred values for true null synapses has a much larger overlap with the distribution of inferred values for truly non-zero synapses).

Among the limitations of the present work, it is worth stressing that we have used integrated-and-fire models with instantaneous synaptic transmission; a natural extension would be to consider models with a more realistic synaptic dynamics. However, we believe we took a step towards making inference in realistic settings, by understanding the role of inherent time scales in neural network dynamics.

\section*{Materials and Methods}
\subsection*{Spiking simulations}
We simulated networks of sparsely connected leaky integrate-and-fire neurons; the $i$-th neuron is modelled as a single dynamic variable, the membrane potential $V_i(t)$, that follows the dynamics:
\begin{equation}
\label{LIF}
\frac{\mathrm{d}V_i(t)}{\mathrm{d}t} = -\frac{V_i(t) - V_{rest}}{\tau_m} + \sum_{j,n} J_{ij} \delta(t - t_{jn} - \delta_{ij}) + J_{ext} \, \sum_n \delta(t - \tilde{t}_{in})
\end{equation}
where $\tau_m = 20$ ms is the membrane's time constant and $V_{rest} = -70$ mV is the membrane's resting potential; $J_{ij}$ measures the (instantaneous, being $\delta(t)$ the Dirac's delta function) post-synaptic potential induced by a spike of neuron $j$ on $V_i$; $t_{jn}$ is the time of firing of the $n$-th spike by neuron $j$, that induces a jump $J_{ij}$ in $V_i$ after a delay $\delta_{ij}$. $J_{ext} = 0.9$ mV models the post-synaptic potential caused by spikes coming from neurons not belonging to the network and collectively firing at a frequency $\nu_{ext} = 1$ kHz; the time $\tilde{t}_{in}$ of arrival of the $n$-th external spike on neuron $i$ is randomly chosen such that $\tilde{t}_{in} - \tilde{t}_{in-1}$ are exponentially distributed with mean $1 / \nu_{ext}$, so that the count of external spikes in a time window $\Delta t$ will follow a Poisson distribution of mean $\nu_{ext} \, \Delta t$. Neuron $i$ will emit a spike at time $t$ whenever $V_i(t^-) \geq \theta = -52$ mV; upon this event, its membrane potential is instantaneously brought to a reset potential which we take equal to the resting value, $V_i(t^+) = V_{rest}$, where it will stay unaffected for a time $\tau_{refractory} = 2$ ms (such condition effectively bounds the firing rate of the neurons below $500$ Hz).

The synaptic efficacies $J_{ij}$ are chosen randomly so that (on average) a fraction $1 - c$ of them is $0$ (that is, neuron $j$ and neuron $i$ are not connected), whereas with probability $c$ they are drawn from a (continuous) probability distribution that depends solely on the excitatory or inhibitory nature of the pre-synaptic neuron $j$. In this paper, we use three different distributions: a delta function (all the synaptic efficacies have equal values), a uniform distribution between two extremal values ($J_{min}$, $J_{max}$), and a Gaussian distribution with given mean $\overline{J}$ and variance $\sigma_J^2$.
When $J_{ij}$ is not zero, a transmission delay $\delta_{ij}$ is sampled from a probability distribution that, in the reported simulations, is either a delta function (all the synapses share a single delay $\overline{\delta}$) or an exponential distribution of mean $\overline{\delta}$ and an offset $\delta_{min}$.
The distribution is truncated at a maximum value $\delta_{max}$: if $\delta_{ij} > \delta_{max}$, then the value is re-sampled ($\delta_{max}$ is chosen so that a re-sampling is triggered on average 5\% of the times). Note that $\overline{\delta}$, in this case, does not coincide with the average delay.

For the results reported in Figs.~\ref{figure7} to \ref{figure9}, the dynamics of the single neuron, Eq.~(\ref{LIF}), is complemented by a synaptic dynamics implementing a form of synaptic short-term depression \cite{tsodyks1997neural}; each synaptic efficacy $J_{ij}$ is replaced by $J_{ij} \, r_{ij}$, where $r_{ij}$ represents the instantaneous fraction of available synaptic resources, and evolves according to:
\begin{equation}
\dot{r}_{ij} = \frac{1 - r_{ij}}{\tau_r} - u\,r_{ij}\,\sum_{n} \delta(t - t_{jn} - \delta_{ij})
\end{equation}
where $\tau_r = 800$ ms and $u = 0.2$; thus each pre-synaptic spike depletes synaptic resources by a fraction $u{}$; synaptic resources, in turn, recover toward $1$ (full availability) with a time constant $\tau_r$. Under the assumption of constant pre-synaptic firing rate $\nu_j$:
\begin{equation}
\label{eq.rSTDStationary}
\langle r_{ij} \rangle_{ss} = \frac{1}{1 + u\,\tau_r\,\nu_j}
\end{equation}
Short-term synaptic depression has been proposed as an activity-dependent network self-inhibition promoting oscillatory or bursting behavior \cite{holcman2006emergence}.

All the simulations have been run using the simulator described in \cite{mattia2000efficient}, that implements an event-driven simulation strategy that does not make use of an integration time-step (that would represent an effective lower-bound for admissible binarization $dt$), allowing to record spike events with arbitrary temporal resolution (within the allowed numerical precision).

\subsection*{Inverse Kinetic Ising Model}
Following \cite{hertz2011ising}, we work with time-binned spike trains under the assumption that, for the chosen time-bin $dt$, there is (almost) never more than one spike per bin. We denote the spike train of neuron $i$ by $S_i(t)$, where $S_i(t) = 1$ if neuron $i$ fired a spike in bin $t$, and $S_i(t) = 0$ otherwise (note that in \cite{hertz2011ising} the convention is, instead, $S_i = \pm 1$). Thus the data we work with is a $N \times T$ binary ``spike matrix'', where $N$ is the total number of neurons considered and $T$ is the number of time-bins. This representation of the data lends itself to formulating the problem in terms of an Ising model, more specifically, given the noisy and evolving nature of spike trains, to a stochastic dynamic formulation of it \cite{glauber1963time, marre2009prediction, roudi2011mean}. At each time-step, for every neuron $i$ in the network we compute the total ``field'':
\begin{equation}
H_{i}(t)=h_{i}+\sum_{j}J_{ij}S_{j}(t)
\end{equation}
where  $h_i$ is an external field. Then we let $S_i$ sample its next value from the probability distribution:
\begin{equation}
P(S_{i}(t + dt)|\mathbf{S}(t))=\frac{\exp{\left[S_{i}(t + dt)H_{i}(t)\right]}}{1+\exp{H_{i}(t)}}
\end{equation}
that depends on the state of the system only at the previous time-step (Markovian dynamics). Thus we are able to compute the likelihood that the probabilistic model generated the binarized data:
\begin{eqnarray}
\emph{L} [S,J,h] &=& \sum_{t = 1}^{T - 1} \sum_{i = 1}^N \ln(P(S_{i}(t + dt)|\mathbf{S}(t)\}) ) = \\
&=&\sum_{it}\bigg[ S_{i}(t + dt) \, H_{i}(t) - \ln \big( 1+\exp(H_{i}(t)) \big) \bigg]
\end{eqnarray}
and, in principle, maximize it to obtain the ``best'' parameters $J$ and $h$. The maximization can be performed iteratively using the gradient:
\begin{eqnarray}
\frac{\partial \emph{L}}{\partial h_i} &=& \langle S_{i}(t)\rangle_{t}- \frac{1}{2} \langle1 + \tanh \frac{H_{i}(t)}{2}\rangle_{t} \nonumber \\
\frac{\partial \emph{L}}{\partial J_{ij}} &=& \langle S_{i}(t + dt) S_{j}(t) \rangle_t-\frac{1}{2} \langle \big(1 + \tanh \frac{H_{i}(t)}{2}\big) \, S_{j}(t) \rangle_{t} \nonumber
\end{eqnarray}
or, in order to avoid computationally expensive iterations and following again \cite{hertz2011ising}, it is possible to use the mean-field equations:
\begin{equation}
\label{campomedio}
m_i=\frac{1}{2}\left[1 + \tanh \left(\frac{h_i+\sum_j J_{ij} m_j}{2}\right)\right]
\end{equation}
where $m_i = \langle S_i \rangle$, and then, assuming that the fluctuations around mean values  $\delta S_i \equiv S_i - m_i$ are small, to write:
\begin{equation}
\label{eq.dLDJMF}
\frac{\partial \emph{L}}{\partial J_{ij}} \simeq D_{ij}(dt) - m_i \, (1 - m_i) \, \sum_{k=1}^N J_{ik} \, C_{kj}
\end{equation} 
where:
\begin{equation}
\label{eq.DIjTau}
D_{ij}(\tau) = \langle \delta S_i(t+\tau) \delta S_j(t) \rangle_{t} \equiv m_j \; (\langle S_i(t + \tau) | S_j(t)=1 \rangle - m_i)
\end{equation}
is the delayed correlation matrix that, for $\tau = 0$, gives the connected correlation matrix:
\begin{equation}
C_{ij} \equiv \langle \delta S_i(t) \delta S_j(t) \rangle_{t} \equiv D_{ij}(0)
\end{equation}
whose diagonal elements are $C_{jj} = m_j \, (1 - m_j)$. Putting $\partial L / \partial J_{ij} = 0$, Eq.~(\ref{eq.dLDJMF}) becomes a set of linear equations that, through a simple matrix inversion, gives the synaptic matrix $J$; then, inverting the (non-linear) equations ~(\ref{campomedio}), also the local static fields $h_i$ are inferred.

\subsection*{Inverse Kinetic Ising Model with time delays}
We extend the Kinetic Ising Model to account for a distribution of delays $\delta_{ij}$ by re-writing the local field as:
\begin{equation}
H_i(t)=h_i+\sum_{j}J_{ij}S_j(t - \delta_{ij})
\label{Hlocale_nuovo}
\end{equation}
This extension does not formally modify the expression for the likelihood of the model on the data, once the new form of $H_i$ is taken into account, whereas the mean-field approximation of the gradient now reads:
\begin{eqnarray}
\label{eq.dLDJMFTD}
&&\frac{\partial \emph{L}}{\partial J_{ij}} \simeq D_{ij}(\delta_{ij}) - m_i \, (1 - m_i) \, \sum_{k=1}^N J_{ik} \, D_{kj}(\delta_{ij} - \delta_{ik}) = \nonumber \\
&=& D_{ij}(\delta_{ij}) - m_i \, (1 - m_i) \, J_{ij} \, C_{jj} - m_i \, (1 - m_i) \, \sum_{k \neq j} J_{ik} \, D_{kj}(\delta_{ij} - \delta_{ik})
\end{eqnarray}
where in the second passage, in the sum over $k$, we singled out the diagonal term that in general will dominate the remainder of the sum, since $D_{kj}(\tau)$, for $k \neq j$, will be typically much smaller than $C_{jj}$ at every $\tau$, and even more so whenever $\tau$ is chosen far from $\delta_{kj}$, where the cross-correlation attains its peak value (see Fig.~\ref{figure3}; here ``far'' is roughly intended with respect to the width of the peak of the cross-correlation); this latter case is the most probable, in the sum in Eq.~(\ref{eq.dLDJMFTD}), since the converse would imply $\delta_{ij} \simeq \delta_{kj} + \delta_{ik}$, an event that is unlikely as long as the the typical values of the delays are larger than the width of the han the $D_{ij}$ peak.

Putting $\partial L / \partial J_{ij} = 0$, and defining a set of $N$ matrices $M^{(i)}$, with $i=1,\ldots,N$, whose elements are
\begin{equation}
M_{kj}^{(i)} \equiv D_{kj}(\delta_{ij} - \delta_{ik})
\end{equation}
Eq.~(\ref{eq.dLDJMFTD}) can be written as $N$ sets, one for each row of the matrix $J$, of $N$ linear equations:
\begin{equation}
D_{ij}(\delta_{ij}) = m_i \, (1 - m_i) \, \sum_{k = 1}^N J_{ik} \, M_{kj}^{(i)}
\end{equation}
that allow, for each post-synaptic neuron $i$, by inverting the matrix $M^{(i)}$, to infer the synaptic couplings with its pre-synaptic neurons. The local fields $h_i$ are found, as above, by then inverting Eq.~(\ref{campomedio}).

\section*{Relationship between synaptic efficacies and inferred couplings}
In order to remap the inferred values onto quantitatively reliable estimates of the synaptic efficacies, we start by putting $\partial L / \partial J_{ij} = 0$ in Eq.~(\ref{eq.dLDJMFTD}) and noting that, by entirely neglecting the typically small (see above) $k \neq j$ terms in the sum, we can approximate the inferred $J_{ij}$ as:
\begin{equation}
\label{eq.jInferApprox}
J_{Inf,ij} \simeq \frac{D_{ij}}{C_{jj} \, m_i \, (1 - m_i)} = \frac{\langle S_i | S_j = 1 \rangle - m_i}{m_i \, (1 - m_i) \, (1 - m_j)}
\end{equation}
where, in the second passage, we made use of Eq.~(\ref{eq.DIjTau}). For reference, having binarized the spike trains with a time-bin $dt$, we can write the average ``magnetization'' $m_i = \nu_i \, dt$, where $\nu_i$ is the average spike-frequency of neuron $i$.
Therefore, we want to estimate the probability that neuron $i$ will fire ($S_i = 1$) upon receiving a spike from neuron $j$ (\textit{i.e.}, conditioned on the event $S_j = 1$).

To start, we note that in Eq.~(\ref{LIF}) the current felt by neuron $i$ in the network can be well approximated, over time scales of the order of the membrane's time constant $\tau_m$, with a Gaussian memoryless stochastic process, identified by a given infinitesimal mean $\mu$ and variance $\sigma^2$. This diffusion approximation is warranted by the small size of the synaptic efficacies and the high frequency of the incoming spikes \cite{renart2004mean}. Under such approximation, Eq.~(\ref{LIF}) describes an Ornstein-Uhlenbeck process with an absorbing barrier \cite{cox1977theory}, whose probability distribution $p(V,\,t)$ obeys the boundary condition:
\begin{equation}
p(\theta,\,t) = 0
\label{absorbing}
\end{equation}
Moreover, it can be shown that the instantaneous emission rate of the neuron is given by:
\begin{equation}
\label{eq.nuAsFlux}
\nu(t) = -\frac{\sigma^2}{2} \, \frac{\partial p(V,\,t)}{\partial V} \Big\vert_{V = \theta}
\end{equation}

Now when a spike from neuron $j$ arrives at neuron $i$ at a (random) time $t + \epsilon$ within the current time-bin $(t,\,t+dt)$ ($0 \leq \epsilon \leq dt$), it produces a sudden jump $J_{ij} \neq 0$ in the membrane potential $V_i{}$; $\epsilon$ is assumed to have uniform probability distribution between $0$ and $dt$, since the only information is the arrival of the spike at some time during $dt$. In the same time-bin $dt$, the neuron also receives external spikes and recurrent spikes from other neurons in the network; in the following, we will assume that both the external and the internal contributions to the input current to the post-synaptic neuron are Poissonian trains; to simplify the discussion, we will assume that all the incoming spikes come through synapses with synaptic efficacy $J_{ext}$ and with a total frequency $\nu_{ext}$: such assumption is easily relaxed to account for pre-synaptic neurons (other than $j$) having different synaptic efficacies $J_{ik}$ and firing frequencies. Finally, in the time interval $\epsilon$ preceding the arrival of the spike from pre-synaptic neuron $j$, we assume that the probability for neuron $i$ to fire is the baseline probability $\nu_i\,\epsilon$; knowing that a pre-synaptic spike from neuron $j$ is due at a given time, indeed, will in general offset this probability, but here we are neglecting this information, assuming the effects are small. The combined contribution of the baseline firing probability, the pre-synaptic spike $S_j = 1$, and the external spikes, in the time-bin $dt$, can be estimated as follows. 

When $J_{ij} > 0$, then instantaneously the whole right tail ($\theta - J_{ij} < V < \theta$) of $p(V,\,t)$ will pass the threshold; in addition, we should consider the contribution from $V < \theta - J_{ij}$ deriving from realizations of the neuron $i$ process that will cross the threshold upon receiving external (excitatory) spikes in the following $(t + \epsilon,\,t+dt)$ interval. For $k$ incoming external spikes, the whole interval of $V$ between $\theta - J_{ij}$ and $\theta - J_{ij} - k\,J_{ext}$ will contribute to the firing probability; an additional element to be taken into account is a drift term due to the leaky dynamics of $V$ Eq.~(\ref{LIF}):
\begin{eqnarray}
&&\langle S_i | S_j = 1 \rangle = \frac{1}{dt} \int_0^{dt} \mathrm{d}\epsilon \, \Big[\nu_i\,\epsilon + \\
&&\sum_{k \geq 0}^\infty \mathrm{Poisson}[k|\nu_{ext}\,(dt - \epsilon)] \, \langle \int_{\theta - k \, J_{ext} - J_{ij} - \mu_0 \, \tilde{t}_k \, (dt - \epsilon)}^{\theta} p(v,\,t) \, \mathrm{d} v \rangle_{\tilde{t}_k} \Big] \simeq \nonumber \\
&\simeq& \frac{\nu_i \, dt}{2} + \frac{1}{dt} \int_0^{dt} \mathrm{d}\epsilon \, \Big[\frac{2\,\nu_i}{\sigma^2} \, \mathrm{Poisson}[0|\nu_{ext}\,(dt - \epsilon)] \, \int_{0}^{J_{ij}} x \, \mathrm{d}x + \nonumber \\
&+& \mathrm{Poisson}[1|\nu_{ext}\,(dt - \epsilon)] \;
\frac{2\,\nu_i}{\sigma^2} \, \langle \int_0^{J_{ij} + J_{ext} + \mu_0\,\tilde{t}_1 \, (dt - \epsilon)} x \, \mathrm{d}x \rangle_{\tilde{t}_1} \Big] \nonumber
\end{eqnarray}
where, in the second passage, we made use of Eq.~(\ref{eq.nuAsFlux}) to approximate $p(V,\,t)$ close to the threshold $\theta$, and neglected the Poisson terms for $k > 1$, that are order $dt^2$ or higher; $\mu_0$ is a constant approximation of the leakage term in the $V$ dynamics (Eq.~(\ref{LIF})) close to the threshold, and $0 \leq \tilde{t}_k \leq 1$ is a random number representing the time of arrival of the $k$-th external spike in a time window (given that exactly $k$ external spikes are due in the time window); it is easy to show that $\langle \tilde{t}_k \rangle = \frac{k}{k + 1}$. Combining this expression with Eq.~(\ref{eq.jInferApprox}) and keeping only the leading terms in $dt$ (averaging over $\tilde{t}_1$ has influence at order $dt^2$ or higher), we have:
\begin{equation}
\label{Jpos1}
J_{Inf,ij} \simeq \frac{J_{ij}^2}{\sigma^2\,dt} + \frac{J_{ij}^2\,(\nu_i + \nu_j) + J_{ij} \, J_{ext} \,\nu_{ext}}{\sigma^2} + O(dt) \quad \mathrm{if} \; J_{ij} > 0
\end{equation}
To derive this expression we furthermore used the equality $\sigma^2 = J_{ext}^2\,\nu_{ext}$ that holds when the input current is comprised of a single Poissonian train of impulses; as stated above, it is easy to relax this assumption and to generalize the result for the case of many trains with different frequencies $\nu_{ext,k}$ and synaptic efficacies $J_{ext,k}$.

Thus, for $J_{ij}>0$, the value of the inferred synaptic coupling $J_{Inf,ij}$ depends quadratically on the real value $J_{ij}$, and critically on $dt$ ($J_{Inf} \sim 1 / dt$); this latter dependence derives from the fact that the contribution to $\langle S_i | S_j = 1 \rangle$ from the spike of pre-synaptic neuron $j$ is order $1$ ($\mathrm{Poisson}[0|\nu_{ext}\,dt] \simeq 1$ for $dt \rightarrow 0$), whereas at the same time, in Eq.~(\ref{eq.jInferApprox}), the denominator vanishes with $dt$ ($m_i \, (1 - m_i) \simeq \nu_i \, dt$).

When $J_{ij} < 0$, the only chance for neuron $i$ to fire after the arrival of a spike from neuron $j$ at time $t + \epsilon$ is exclusively given by the probability that one or more external spikes will compensate for the sudden negative jump $J_{ij}$, making $V_i$ pass the threshold $\theta$:
\begin{eqnarray}
\label{eq.poissonExpansion}
&& \langle S_i | S_j = 1 \rangle = \frac{1}{dt} \int_0^{dt} \mathrm{d}\epsilon \, \Big[\nu_i\,\epsilon + \\
&& \sum_{k \geq 1}^\infty \mathrm{Poisson}(k|\nu_{ext}\,(dt - \epsilon)) \; \langle \int_{\theta - k \, J_{ext} + |J_{ij}| - \mu_0 \, \tilde{t}_k \, (dt - \epsilon)}^{\theta} p(v,\,t) \, \mathrm{d} v \rangle_{\tilde{t}_k} \Big] \nonumber
\end{eqnarray}
It is important to note that, if $|J_{ij}| > J_{ext}$, the term with $k = 1$ will vanish (a single external spike won't suffice to compensate for the negative jump $J_{ij}$); if $|J_{ij}| > 2 \, J_{ext}$, the second term too will disappear (not even two external spikes will be enough) and so on.

Now let's assume that $|J_{ij}| < J_{ext}$, so that all the terms in the sum are non-zero and the sum will be dominated, for small $dt$, by the first term only; then, reasoning as above for $J_{ij} > 0$, we have:
\begin{equation}
\label{Jneg1}
J_{Inf,ij} \simeq \frac{(J_{ij}^2 + 2\,J_{ij}\,J_{ext}) \, \nu_{ext}}{2\,\sigma^2} + O(dt) \quad \mathrm{if} \; -J_{ext} \leq J_{ij} < 0
\end{equation}
Under the hypothesis $|J_{ij}| < J_{ext}$, then, the dependence of the inferred coupling on the real synaptic efficacy is again quadratic, as for the $J_{ij} > 0$ case, but its leading term does not depend on $dt$.

If we assume, instead, $J_{ext} < |J_{ij}| \leq 2 \, J_{ext}$, the sum in Eq.~(\ref{eq.poissonExpansion}) will be dominated by the term $k = 2$, since the first term vanishes, and thus we get:
\begin{equation}
\label{eq.jInferInhGreaterJExt}
J_{Inf,ij} \simeq -\frac{1}{2} + O(dt) \quad \mathrm{if} \; -2\,J_{ext} \leq J_{ij} < -J_{ext}
\end{equation}
Basically then, for $|J_{ij}| > J_{ext}$, the inferred $J$ will be largely independent of the true value of $J_{ij}$. This result can be intuitively understood by examining the extreme case $J_{ij} \rightarrow -\infty$; in this case, the only surviving contribution to the probability of firing during $dt$ (given the arrival of the inhibitory spike from $j$), is the baseline probability $\nu_i\,\epsilon$ before the large, negative jump $J_{ij}$ in $V_i$; such term, integrated over a uniform distribution becomes $\nu_i \, dt / 2$, which inserted into Eq.~(\ref{eq.jInferApprox}) gives $J_{Inf,ij} \simeq -1/2$. Eq.~(\ref{eq.jInferInhGreaterJExt}) then shows that, for $dt \rightarrow 0$, the limit case's behavior of $J_{Inf}$ is essentially attained already for $J_{ij} \leq -J_{ext}$.

\section*{Acknowledgments}
We thank Y. Roudi for interesting discussions.


\begin{thebibliography}{10}
\providecommand{\url}[1]{\texttt{#1}}
\providecommand{\urlprefix}{URL }
\expandafter\ifx\csname urlstyle\endcsname\relax
  \providecommand{\doi}[1]{doi:\discretionary{}{}{}#1}\else
  \providecommand{\doi}{doi:\discretionary{}{}{}\begingroup
  \urlstyle{rm}\Url}\fi
\providecommand{\bibAnnoteFile}[1]{%
  \IfFileExists{#1}{\begin{quotation}\noindent\textsc{Key:} #1\\
  \textsc{Annotation:}\ \input{#1}\end{quotation}}{}}
\providecommand{\bibAnnote}[2]{%
  \begin{quotation}\noindent\textsc{Key:} #1\\
  \textsc{Annotation:}\ #2\end{quotation}}
\providecommand{\eprint}[2][]{\url{#2}}

\bibitem{eggermont1990correlative}
Eggermont JJ (1990) The correlative brain.
\newblock Springer.
\input{eggermont1990correlative}

\bibitem{pernice2013reconstruction}
Pernice V, Rotter S (2013) Reconstruction of sparse connectivity in neural
  networks from spike train covariances.
\newblock Journal of Statistical Mechanics: Theory and Experiment 2013: P03008.
\input{pernice2013reconstruction}

\bibitem{bialek2006weak}
Schneidman E, Berry MJ, Segev R, Bialek W (2006) Weak pairwise correlations
  imply strongly correlated network states in a neural population.
\newblock Nature 440: 1007--1012.
\input{bialek2006weak}

\bibitem{marre2009prediction}
Marre O, El~Boustani S, Fr{\'e}gnac Y, Destexhe A (2009) Prediction of
  spatiotemporal patterns of neural activity from pairwise correlations.
\newblock Physical Review Letters 102: 138101.
\input{marre2009prediction}

\bibitem{hertz2011ising}
Hertz J, Roudi Y, Tyrcha J (2013) Ising models for inferring network structure
  from spike data.
\newblock In: Quian~Quiroga R, Panzeri S, editors, Principles of neural coding,
  CRC Press.
\input{hertz2011ising}

\bibitem{roudi2011mean}
Roudi Y, Hertz J (2011) Mean field theory for nonequilibrium network
  reconstruction.
\newblock Physical Review Letters 106: 048702.
\input{roudi2011mean}

\bibitem{bialek2012biophysics}
Bialek W (2012) Biophysics: searching for principles.
\newblock Princeton University Press.
\input{bialek2012biophysics}

\bibitem{truccolo2005point}
Truccolo W, Eden UT, Fellows MR, Donoghue JP, Brown EN (2005) A point process
  framework for relating neural spiking activity to spiking history, neural
  ensemble, and extrinsic covariate effects.
\newblock Journal of neurophysiology 93: 1074--1089.
\input{truccolo2005point}

\bibitem{pillow2008spatio}
Pillow JW, Shlens J, Paninski L, Sher A, Litke AM, et~al. (2008)
  Spatio-temporal correlations and visual signalling in a complete neuronal
  population.
\newblock Nature 454: 995--999.
\input{pillow2008spatio}

\bibitem{roudi2015multi}
Roudi Y, Dunn B, Hertz J (2015) Multi-neuronal activity and functional
  connectivity in cell assemblies.
\newblock Current opinion in neurobiology 32: 38--44.
\input{roudi2015multi}

\bibitem{glauber1963time}
Glauber RJ (1963) Time-dependent statistics of the ising model.
\newblock Journal of Mathematical Physics 4: 294--307.
\input{glauber1963time}

\bibitem{renart2004mean}
Renart A, Brunel N, Wang XJ (2004) Mean-field theory of irregularly spiking
  neuronal populations and working memory in recurrent cortical networks.
\newblock Computational Neuroscience: A comprehensive approach : 431--490.
\input{renart2004mean}

\bibitem{tsodyks1997neural}
Tsodyks MV, Markram H (1997) The neural code between neocortical pyramidal
  neurons depends on neurotransmitter release probability.
\newblock Proceedings of the National Academy of Sciences 94: 719--723.
\input{tsodyks1997neural}

\bibitem{holcman2006emergence}
Holcman D, Tsodyks M (2006) The emergence of up and down states in cortical
  networks.
\newblock PLoS Computational Biology 2: e23.
\input{holcman2006emergence}

\bibitem{eytan2006dynamics}
Eytan D, Marom S (2006) Dynamics and effective topology underlying
  synchronization in networks of cortical neurons.
\newblock The Journal of Neuroscience 26: 8465--8476.
\input{eytan2006dynamics}

\bibitem{roudi2009pairwise}
Roudi Y, Nirenberg S, Latham PE (2009) Pairwise maximum entropy models for
  studying large biological systems: when they can work and when they can't.
\newblock PLoS Computational Biology 5: e1000380.
\input{roudi2009pairwise}

\bibitem{roudi2009statistical}
Roudi Y, Aurell E, Hertz JA (2009) Statistical physics of pairwise probability
  models.
\newblock Frontiers in Computational Neuroscience 3.
\input{roudi2009statistical}

\bibitem{roudi2009}
Roudi Y, Tyrcha J, Hertz J (2009) Ising model for neural data: Model quality
  and approximate methods for extracting functional connectivity.
\newblock Physical Review E 79: 051915.
\input{roudi2009}

\bibitem{cocco2009neuronal}
Cocco S, Leibler S, Monasson R (2009) Neuronal couplings between retinal
  ganglion cells inferred by efficient inverse statistical physics methods.
\newblock Proceedings of the National Academy of Sciences 106: 14058--14062.
\input{cocco2009neuronal}

\bibitem{mattia2000efficient}
Mattia M, Del~Giudice P (2000) Efficient event-driven simulation of large
  networks of spiking neurons and dynamical synapses.
\newblock Neural Computation 12: 2305--2329.
\input{mattia2000efficient}

\bibitem{cox1977theory}
Cox DR, Miller HD (1965) The theory of stochastic processes.
\newblock Chapman \& Hall.
\input{cox1977theory}

\end{thebibliography}
\end{document}